\documentclass[final,5p,times,twocolumn]{elsarticle}
\usepackage{algorithmic}
\usepackage{textcomp}
\usepackage{amsmath,amssymb,amsfonts}
\usepackage{todonotes}
\usepackage{circuitikz}
\usepackage{color}
\usepackage{tikz}
\usepackage{hyperref}
\usepackage{mathtools}
\usepackage{wrapfig}
\usepackage[mathlines]{lineno}

\newcommand\EqFontSize{\large}

\tikzset{
  declare function={
    atan3(\a,\b)=ifthenelse(atan2(0,1)==90, atan2(\a,\b), atan2(\b,\a));},
  kinky cross radius/.initial=+.125cm,
  @kinky cross/.initial=+, kinky crosses/.is choice,
  kinky crosses/left/.style={@kinky cross=-},kinky crosses/right/.style={@kinky cross=+},
  kinky cross/.style args={(#1)--(#2)}{
    to path={
      let \p{@kc@}=($(\tikztotarget)-(\tikztostart)$),
          \n{@kc@}={atan3(\p{@kc@})+180} in
      -- ($(intersection of \tikztostart--{\tikztotarget} and #1--#2)!%
             \pgfkeysvalueof{/tikz/kinky cross radius}!(\tikztostart)$)
      arc [ radius     =\pgfkeysvalueof{/tikz/kinky cross radius},
            start angle=\n{@kc@},
            delta angle=\pgfkeysvalueof{/tikz/@kinky cross}180 ]
      -- (\tikztotarget)}}}
      
\ctikzset{tripoles/mos style=arrows}
\ctikzset{transistors/arrow pos=end}

\journal{Sensors and Actuators A: Physical}
\begin{document}

\begin{frontmatter}

\title{Low-Power Silicon Strain Sensor Based on CMOS Current Reference Topology}
\author[inst1]{Nicolas~Roisin}
\ead{nicolas.roisin@uclouvain.be}

\author[inst1]{Thibault~P.~Delhaye}
\author[inst1]{Nicolas~André}
\author[inst1]{Jean-Pierre~Raskin}
\author[inst1]{Denis~Flandre}

\address[inst1]{Institute of Information and Communication Technologies, Electronics and Applied Mathematics (ICTEAM), UCLouvain, Place du levant 3, 1348 Louvain-la-Neuve, Belgium}
            

\begin{abstract}
A strain sensor inspired by a Widlar self-biased current source topology called $\beta$-multiplier is developed to obtain a strain-dependent reference current with high supply rejection. The sensor relies on the piezoresistive effect in  the silicon MOS transistors that form the current reference circuit. The device behavior is analytically computed and verified with experimental measurements under four-point bending test. A basic implementation with an integrated resistor reaches a strain sensitivity of 2.54 nA/$\mu\epsilon$ (gauge factor of 324) for a temperature sensitivity of 52.06 nA/°C. A more advanced full-transistor circuit based on current subtraction principle is furthered implemented in order to reach strain sensitivity up to 12.02 nA/$\mu\epsilon$ (gauge factor of 1773) and temperature sensitivity of -28.72 nA/°C. 
This implementation includes a CMOS active load to tune the strain and temperature sensitivities with a total power consumption between 20 and 150 $\mu$W.

\end{abstract}



\end{frontmatter}

\section{Introduction}

Many physical effects can be exploited to record strain in a material such as capacitive \cite{Cao2015,Downey2016}, piezoelectric for dynamic strain \cite{Chew2017a,Yamashita2016}, optic with interferometers  \cite{Liehr2012,Guo2016,Song2006} and piezoresistive \cite{Fiorillo2018,Gridchin1995,Zymelka2017a}.
The two last means are widely used, especially with fiber Bragg grating and interferometric sensors for optical measurements and metallic strain gauges for piezoresistive sensing. 
Optical sensors offer various advantages such as accuracy and the ease of multiplexing but comes with brittleness and high power consumption due to the spectrum analyzer that can consume several Watts. Their robustness against external conditions (electromagnetic interferences, chemicals, temperature, and so on.) make them suitable for harsh environments found in industrial areas, biomedical, automotive or aerospace applications \cite{Pettinato2021}.
Piezoresistive strain gauges provide a cheap solution easy to implement. It is based on a mature technology that was one of the first used to record strain of a material \cite{Zhang2019}. The principle of the metallic gauge is based on a change in the dimensions that limits the gauge factor (GF) at around 2. 

New types of materials such as ceramic or semiconductor can be used in order to improve this factor and reach higher strain sensitivity \cite{Jabir2013,Delhaye2021}. The principle of strain sensing with those materials does not rely on dimensions variation but on intrinsic changes in the carrier mobility.
Some of them, such as graphene \cite{Zheng2020} or carbon nanotubes \cite{Ke2018}, allow high gauge factor up to 1000. However, the fabrication cost and complexity, added to the fragility, make long-term applications difficult.
On the other hand, silicon is a well-known material in electronics with reasonable cost and fabrication complexity \cite{Fisher2012}. Its crystallographic configuration leads to high piezoresistive variation with potential gauge factor above 150 \cite{Taroni1970}.

One of the biggest advantage of silicon gauges lies in its straightforward integration in a complete circuit to boost and tune the performances of the sensor. 
In this case, the sensing element is no more an external component which has to be inserted in a dedicated circuit, but the circuit itself.

Self-biased reference circuit refer to implementations where the reference current or voltage is set independently of external sources \cite{Vittoz1977}. They provide stable output signal that depends on physical parameter such as carrier mobility, resistance, breakdown voltage, dimensions and so on. In this work, a self-biased current reference inspired by Widlar topology, called $\beta$-multiplier \cite{Song1998}, is modified in order to create a strain-dependent reference current with high supply rejection.
Four implementations are analyzed: (i) a basic implementation with a resistor ($\beta_R$), (ii) a full-transistor implementation with positive ($\beta_+$) or (iii) negative ($\beta_-$) strain response and (iv) a current subtraction implementation based on full-transistor circuits ($\beta_{sub}$).

The analytical analysis of the circuits is first realized in Section \ref{sec:analytical} to present the different implementations and to model the strain impact on the reference currents.
Then, the results of measurements on the sensor are presented and discussed in Section \ref{sec:experimental}. We first tested the transistors separately in order to extract the key parameters, i.e. the carrier mobility and the threshold voltage, along with their strain and temperature dependencies. We complete the study with the strain and temperature analyses of the $\beta$-multipliers to evaluate the performances of the different implementations.

\section{Analytical Model}
\label{sec:analytical}

\subsection{Piezoresistive effect}

The piezoresistive effect in silicon is mainly due to the change in the carrier mobility that is much higher than the change in the dimensions \cite{johns2006}. By considering infinitesimal displacement, the relative variation of the mobility can be expressed as 
\begin{equation}\label{eq:linear}\EqFontSize
    -\frac{d\mu_i}{\mu_i}=\pi_{i,l}\sigma_l+\pi_{i,t}\sigma_t,
\end{equation}
where $i$ stands for the electron (n) or holes (p) contributions, $\pi_l$ and $\pi_t$ are, respectively, the piezoresistive coefficients in the longitudinal and transverse directions while $\sigma_l$ and $\sigma_t$ are the applied stresses in those two directions. We consider the longitudinal direction to be the direction of the transistor channel while the transverse direction is perpendicular to it. 
In ceramic material, the strain-stress relation can be expressed by considering elastic deformation \cite{Jastrzebski1988} as
\begin{equation}\EqFontSize
    \sigma_j=E\cdot\varepsilon_j,
\end{equation}
where $E$ is the Young's modulus of around 165 GPa for silicon in the [110] crystal direction \cite{Hopcroft2010} and $\varepsilon_j$ is the applied strain in direction $j$.

Solving relation \ref{eq:linear} leads to
\begin{equation}\EqFontSize\label{eq:muexp}
    \mu_i=\mu_{i,0}~ e^{-(\pi_{i,l} \sigma_l+\pi_{i,l}\sigma_t)},
\end{equation}
where $\mu_{i,0}$ is the mobility in the relaxed case. 

The piezoresistive coefficients are intrinsic parameters that are linked to the effective masses variations of the electrons and the holes. As the variations of the effective mass depend on the crystal direction of the applied strain, so do these coefficients \cite{Tan2008}. 

\begin{figure}[htb!]
    \centering
    \resizebox{9cm}{!}{
\begin{tikzpicture}
    \node (image) at (0,0) {
        \includegraphics[width=0.3\textwidth]{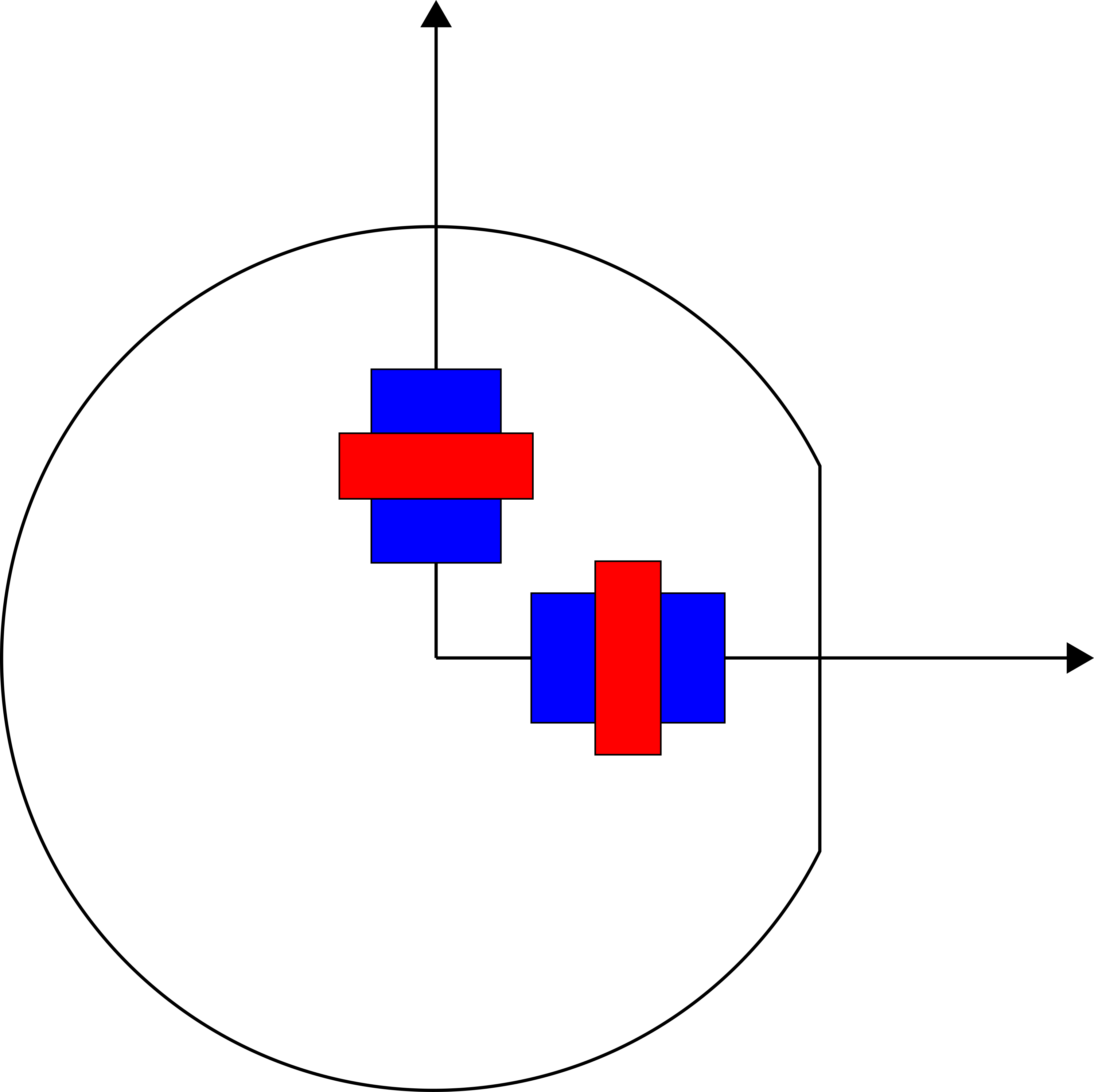}
    };
    \draw[thick,-Latex,gray] (0,-1.2) -- (1,-1.2) node[midway, anchor=north]{$J$};
    \draw[thick,-Latex,gray] (-1.2,0) -- (-1.2,1) node[midway, anchor=east]{$J$};
    \draw (-0.3,2.7) node[] {$y$};
    \draw (-0.5,3.2) node[] {[$\bar{1}10$]};
    \draw (2.7,-0.3) node[] {$x$};
    \draw (3.3,-0.5) node[] {[$110$]};
    \draw (-1.4,-1.4) node[] {(100) wafer};
\end{tikzpicture}   
}
\caption{Illustration of the transistors orientations along $<$110$>$ directions on a (100) silicon wafer.}
    \label{fig:sketch}
\end{figure}
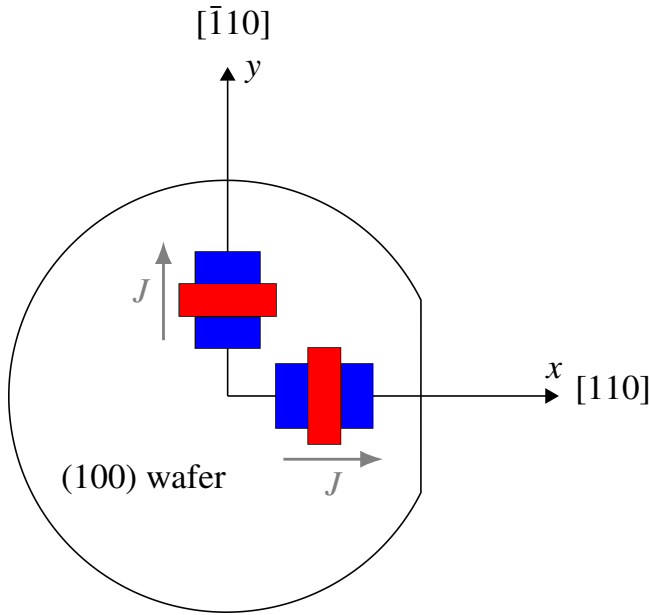

In this work, we consider MOS transistor with channel oriented in parallel and perpendicular to the [110] direction of the crystal as displayed in Fig. \ref{fig:sketch}. 

In this orientation, the transverse and longitudinal piezoresistive coefficients $\pi_{i,l}$ and $\pi_{i,t}$ can be expressed as \cite{Kanda1982}
\begin{equation}\EqFontSize
    \begin{array}{ccl}
         \pi_{i,t}&=&\frac{\pi_{i,11}+\pi_{i,12}-\pi_{i,44}}{2},  \\&&\\
         \pi_{i,l}&=&\frac{\pi_{i,11}+\pi_{i,12}+\pi_{i,44}}{2},  \\
    \end{array}
\end{equation}
where $\pi_{i,11}$, $\pi_{i,12}$ and $\pi_{i,44}$ are the main components of the piezoresistance tensor. The values for n-type and p-type silicon  are given in Table \ref{tab:piezoSi} \cite{Smith1954}.

\begin{table}[hbpt]
    \centering
    \renewcommand{\arraystretch}{1.3}
    \begin{tabular}{c|ccc|cc}
    \hline
       &$\pi_{11}$&$\pi_{12}$&$\pi_{44}$&$\pi_{t}$&$\pi_{l}$\\\hline
       n-Si  & -1022&534&-136&-176&-312 \\
       p-Si  & 66&-11&1381&-663&718\\\hline
    \end{tabular}  
    \vspace{0.2cm}
    \caption{Main components of the piezoresistance tensor for p- and n-type transistors \cite{Smith1954}. The effective longitudinal and transverse piezoresistive coefficients for strain applied in the [110] direction are also computed. The values are expressed in TPa$^{-1}$.}
    \label{tab:piezoSi}
\end{table}

At the circuit level, it was found that perpendicular transistors placed in current mirror configuration lead to higher strain sensitivity with regards to ratio of the transistor currents \cite{Rue2011, Jaeger1995,Jaeger2000}. Indeed, this configuration leads to an effective sensitivity that is equal to the difference of the piezoresistive coefficients of the two transistors. If these coefficients are of opposite sign, an enhanced strain sensitivity can be obtained. Such features can be found by placing PMOS (NMOS) transistors perpendicularly along the [110] ([100]) directions. This perpendicular current mirror works as the main sensing element of our proposed solution for strain sensing applications.

\subsection{Current reference configuration}
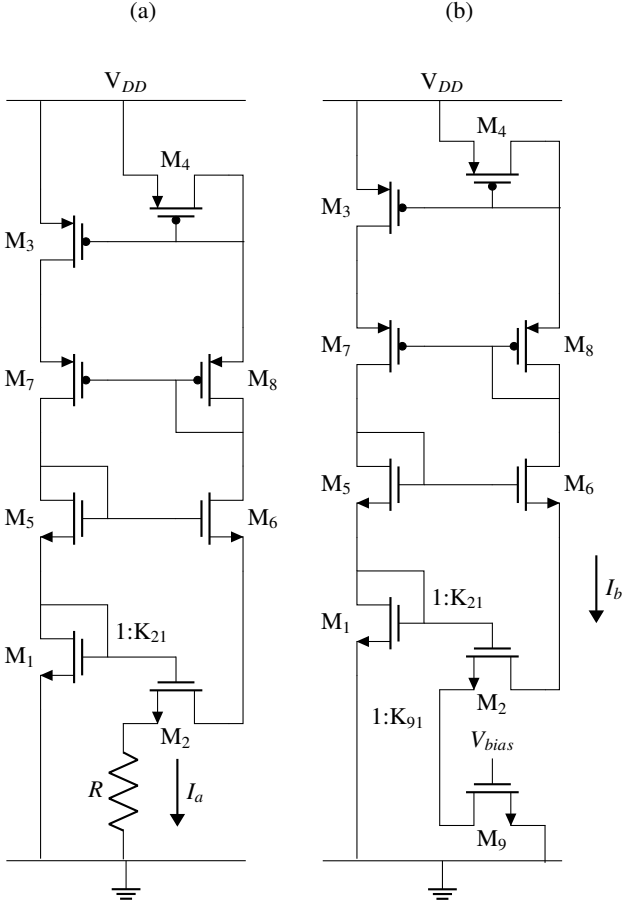
\begin{figure}[!ht]
    \centering
    \begin{minipage}[c]{.46\linewidth}
    \begin{circuitikz}[scale=0.9,transform shape]
\draw (0,0) node[nmos,xscale=-1] (M1){\ctikzflipx{M$_1$}};
\draw (M1.D)--++(0,0.5) node[nmos,xscale=-1,anchor=south east] (M5){\ctikzflipx{M$_5$}};
\draw (M5.D)--++(0,0.5)
node[pmos,xscale=-1,anchor=south east](M7){\ctikzflipx{M$_7$}};
\draw (M7.S)--++(0,0.5) node[pmos,xscale=-1,anchor=south east](M3){\ctikzflipx{M$_3$}};

\draw (M3.G)--++(1,0) node[pmos,anchor=west,rotate=90](M4){\rotatebox[origin=c]{-90}{M$_4$}};
\draw (M7.G)--++(1,0) node[pmos,anchor=west](M8){M$_8$};
\draw (M1.G)to[short,l=1:K$_{21}$]++(1,0) node[nmos,anchor=west,rotate=-90](M2){\rotatebox[origin=c]{90}{M$_2$}};
\draw (M5.G)to[short]++(1,0) node[nmos,anchor=west](M6){M$_6$};

\draw (M2.S) to[R,l_=$R$] ++(0,-2) ;
\draw (M1.S) to[short] ++(0,-2.2) ;

\draw (M3.S) to[short] ++(0,1.3);
\draw (M4.S) to[short] ++(0,1.1);

\draw (M2.D)-|(M6.S);
\draw (M6.D)--(M8.D);
\draw (M8.S)|-(M4.D);
\draw (M4.G)-|++(1,0);
\draw (M1.G)|-(M1.D);
\draw (M8.G)|-(M8.D);
\draw (M5.G)|-(M5.D);

\draw (1.5,9.5) node[] {(a)};
\draw[thick,-Triangle] (2,-1.5) -- (2,-2.5) node[midway,right]{$I_a$};

\draw (-0.5,8.2) -- (3,8.2) node[midway,above] {V$_{DD}$} ;
\draw (-0.5,-3) -- (3,-3) node[ground,midway]{};

\end{circuitikz}\end{minipage}\hfill
    \begin{minipage}[c]{.53\linewidth}
    \begin{circuitikz}[scale=0.9, transform shape]

\draw (0,0) node[nmos,xscale=-1] (M1){\ctikzflipx{M$_1$}};
\draw (M1.D)--++(0,0.5) node[nmos,xscale=-1,anchor=south east] (M5){\ctikzflipx{M$_5$}};
\draw (M5.D)--++(0,0.5)
node[pmos,xscale=-1,anchor=south east](M7){\ctikzflipx{M$_7$}};
\draw (M7.S)--++(0,0.5) node[pmos,xscale=-1,anchor=south east](M3){\ctikzflipx{M$_3$}};

\draw (M3.G)--++(1,0) node[pmos,anchor=west,rotate=90](M4){\rotatebox[origin=c]{-90}{M$_4$}};
\draw (M7.G)--++(1,0) node[pmos,anchor=west](M8){M$_8$};
\draw (M1.G)to[short,l=1:K$_{21}$]++(1,0) node[nmos,anchor=west,rotate=-90](M2){\rotatebox[origin=c]{90}{M$_2$}};
\draw (M5.G)to[short]++(1,0) node[nmos,anchor=west](M6){M$_6$};
\draw (M2.S)--++(0,-2) node[nmos,xscale=-1,rotate=-90,anchor=north east](M9){\rotatebox[origin=c]{90}{\ctikzflipx{M$_9$}}}
(M9.G) node[anchor=south] {$V_{bias}$};

\draw (M1.S) to[short] ++(0,-2.7);
\draw (M9.S) to[short] ++(0,-0.55);
\draw (M2.S) ++(0,-1.75);
\draw (M3.S) to[short] ++(0,0.8);
\draw (M4.S) to[short] ++(0,0.6);

\draw (M2.D)-|(M6.S);
\draw (M6.D)--(M8.D);
\draw (M8.S)|-(M4.D);
\draw (M4.G)-|++(1,0);
\draw (M1.G)|-(M1.D);
\draw (M8.G)|-(M8.D);
\draw (M5.G)|-(M5.D);

\draw (1.5,9) node[] {(b)};
\draw[thick,-Triangle] (3.5,1) -- ++(0,-1) node[midway,right]{$I_b$};
\draw (0.6,-1.4) node[] {1:K$_{91}$};

\draw (-0.5,7.7) -- (3,7.7) node[midway,above] {V$_{DD}$} ;
\draw (-0.5,-3.5) -- (3,-3.5) node[ground,midway]{};

\end{circuitikz}
    \end{minipage}
    \caption{$\beta$-multiplier reference circuits with (a) resistor and (b) full-transistor implementations. The reference currents are, respectively, written $I_a$ and $I_b$.}
    \label{fig:betamult}
\end{figure}

In this work, we first imagined a $\beta$-multiplier-like reference shown in Fig. \ref{fig:betamult}.a as strain sensor. This is inspired by a self-biased reference based on Widlar current source topology. A cascode configuration is implemented in order to further reduce the supply sensitivity.

\begin{figure*}[ht!]
    \centering
    \begin{circuitikz}[thick,scale=1, transform shape]

\draw[ultra thick] (0,-7.6) -- (16.5,-7.6);
\draw (8.25,-7.6) node[ground]{};
\draw[ultra thick] (0,1.5) -- (16.5,1.5) node[midway,above] {\textbf{V}$_{DD}$} ;




\draw (1,0) node[pmos,xscale=-1] (p1) {};
\draw (3,1) node[pmos,rotate=90] (p2) {};
\draw (1,-1.2) node[pmos,xscale=-1] (p3) {};
\draw (4,-1.2)  node[pmos ] (p4) {};

\draw (1,-4.2) node[nmos,xscale=-1] (n1) {};
\draw (3,-5.2) node[nmos,rotate=-90] (n2) {};
\draw (1,-3) node[nmos,xscale=-1] (n3) {};
\draw (4,-3) node[nmos] (n4) {};
\draw (3,-7) node[nmos,xscale=-1,rotate=-90] (n5) {};

\draw (p1.G) node[above] {{\footnotesize M$_3$}};
\draw (p2.G) node[above=9mm] {{\footnotesize M$_4$}};
\draw (n1.G) node[left=9mm] {{\footnotesize M$_1$}};
\draw (n2.G) node[below=10mm] {{\footnotesize M$_2$,K$_{21}$}};
\draw (p3.G) node[left=8mm] {{\footnotesize M$_7$}};
\draw (p4.G) node[right=8mm] {{\footnotesize M$_8$}};
\draw (n3.G) node[left=8mm] {{\footnotesize M$_5$}};
\draw (n4.G) node[above] {{\footnotesize M$_6$}};
\draw (n5.G)--($(n5.G)+(0.5,0)$)  node[right] {{\footnotesize \textbf{V}$_{Bias}$}};
\draw (n5.G) node[below=9mm] {{\footnotesize M$_9$,K$_{91}$}};

\draw (p1.S)-- (1,1.5) -| (p2.S);
\draw (p1.D) -- (p3.S);
\draw (p2.D) -| (p4.S);
\draw (p3.G) -- (p4.G);
\draw (p4.G) |- (p4.D);
\draw (p1.G) -| (p2.G);
\draw (p1.G) -| (p4.S);
\draw (p3.D) -- (n3.D);
\draw (p4.D) -- (n4.D);
\draw (n3.G) -- (n4.G);
\draw (n3.G) |- (n3.D);
\draw (n1.G) -| (n2.G);
\draw (n1.S) |- (3,-7.6);
\draw (n5.S) -| (4,-7.6);
\draw (n5.D) |- (n2.S);
\draw (n1.D)-| (n1.G);

\draw (6,0) node [pmos, xscale=-1] (b1) {};
\draw (6,-1.2) node [pmos, xscale=-1] (b2){};
\draw (9,0) node [pmos] (b3) {};
\draw (9,-1.2) node [pmos] (b4) {};
\draw (10,0) node [pmos, xscale=-1] (b7) {};
\draw (10,-1.2) node [pmos, xscale=-1] (b8){};
\draw (9,-3) node [nmos] (b9) {};
\draw (9,-4.2) node [nmos] (b10) {};
\draw (b1.S) -- (6,1.5);
\draw (b3.S) -- (9,1.5);
\draw (b3.G) |- (b3.D);
\draw (b3.G) -- (b1.G);
\draw (b4.G) |- (b4.D);
\draw (b4.G) -- (b2.G);

\draw (b9.D) -- (b4.D);
\draw (b10.S) -- (9,-7.6);
\draw (b9.G) -- (n4.G);

\draw (n2.D)-|(n4.S) ;
\draw (4,-5.4) node[] (x) {};
\draw (b10.G) to [kinky cross=(x)--(n4.S), kinky crosses=right] (n1.G);

\draw (b7.S) -- (10,1.5);
\draw (b8.D) |- (b9.D);
\draw (b1.G) node[left=8mm] {\footnotesize M$_{22}$};
\draw (b2.G) node[left=8mm] {\footnotesize M$_{23}$};
\draw (b3.G) node[right=8mm] {\footnotesize M$_{25}$};
\draw (b4.G) node[right=8mm] {\footnotesize M$_{26}$};
\draw (b7.G) node[above] {\footnotesize M$_{27}$};
\draw (b8.G) node[above] {\footnotesize M$_{28}$};
\draw (b9.G) node[right=8mm] {\footnotesize M$_{29}$};
\draw (b10.G) node[right=8mm] {\footnotesize M$_{30}$};


\draw (13,1) node[pmos,rotate=-90,xscale=-1] (c1) {};
\draw (15,0) node[pmos] (c2) {};
\draw (12,-1.2) node[pmos,xscale=-1] (c3) {};
\draw (15,-1.2)  node[pmos ] (c4) {};

\draw (13,-5.2) node[nmos,rotate=-90] (c5) {};
\draw (15,-4.2) node[nmos] (c6) {};
\draw (12,-3) node[nmos,xscale=-1] (c7) {};
\draw (15,-3) node[nmos] (c8) {};
\draw (15,-6.6) node[nmos,xscale=-1] (c9) {};

\draw (c1.G) |- (c2.G) |-(c2.D);
\draw (c1.S)|- (14,1.5)-| (c2.S);
\draw (c3.G)--(c4.G)|-(c4.D);
\draw (c4.D)--(c8.D);
\draw (c8.S)--(c6.D);
\draw (c6.S)--(c9.D);
\draw (c6.G) -| (c5.G);
\draw (c3.D)--(c7.D);
\draw (c7.D)-|(c7.G)--(c8.G);
\draw (c6.G)-|(c7.S);
\draw (c5.S) -| (c7.S);
\draw (c5.D) |- (14,-7.6);
\draw (c9.D) -- (c6.S);
\draw (c9.G)-- ($(c9.G)+(0,0.5)$) node [above] {{\footnotesize \textbf{V}$_{Bias}$}};
\draw (c3.G)--(b8.G);

\draw (c1.D)-|(c3.S);
\draw (12,1) node[] (x1){};
\draw (12,-1) node[] (x2){};

\draw (b7.G) to [kinky cross=(x1)--(x2), kinky crosses=left] (c2.G);
\draw (c9.S)|-(15,-7.6);

\draw (c1.G) node[above=9mm] {{\footnotesize M$_{12}$}};
\draw (c2.G) node[above] {{\footnotesize M$_{13}$}};
\draw (c3.G) node[below] {{\footnotesize M$_{16}$}};
\draw (c4.G) node[right=8mm] {{\footnotesize M$_{17}$}};
\draw (c5.G) node[below = 9 mm] {{\footnotesize M$_{10}$}};
\draw (c6.S) node[right] {{\footnotesize M$_{11}$,K$_{21}$}};
\draw (c7.G) node[left=8mm] {{\footnotesize M$_{14}$}};
\draw (c8.G) node[right=8mm] {{\footnotesize M$_{15}$}};
\draw (c9.G) node[left=7mm] {{\footnotesize M$_{18}$,K$_{91}$}};



\draw[thick,-Triangle] (6,-2.1) -- ++(0,-0.7) node[midway,right]{$I_c$};
\draw[thick,-Triangle] (4.3,-4.6) -- ++(0,-0.7) node[midway,right]{$I_{b,+}$};
\draw[thick,-Triangle] (15.3,-3.4) -- ++(0,-0.7) node[midway,right]{$I_{b,-}$};

\draw [draw opacity=0.2, fill=blue, fill opacity=0.05](-0.5,3) -- (-0.5,-9) -- (5,-9) -- (5,3) -- cycle;
\draw [draw opacity=0.2, fill=red, fill opacity=0.05](5,3) -- (5,-9) -- (11.5,-9) -- (11.5,3) -- cycle;
\draw [draw opacity=0.2, fill=blue, fill opacity=0.05](11.5,3) -- (11.5,-9) -- (17,-9) -- (17,3) -- cycle;
\draw[blue,thick] (3,2.5) node[left] {\LARGE $\beta_+$};
\draw[red,thick] (11,2.5) node[left] {\LARGE Current subtraction};
\draw[blue,thick] (15,2.5) node[left] {\LARGE$\beta_-$};
\end{circuitikz}
    \caption{Current subtraction implementation ($\beta_{sub}$) composed full-transistor $\beta$-multipliers with positive ($\beta_{+}$) and negative ($\beta_{-}$) strain response}
    \label{fig:AddCircuit}
\end{figure*}
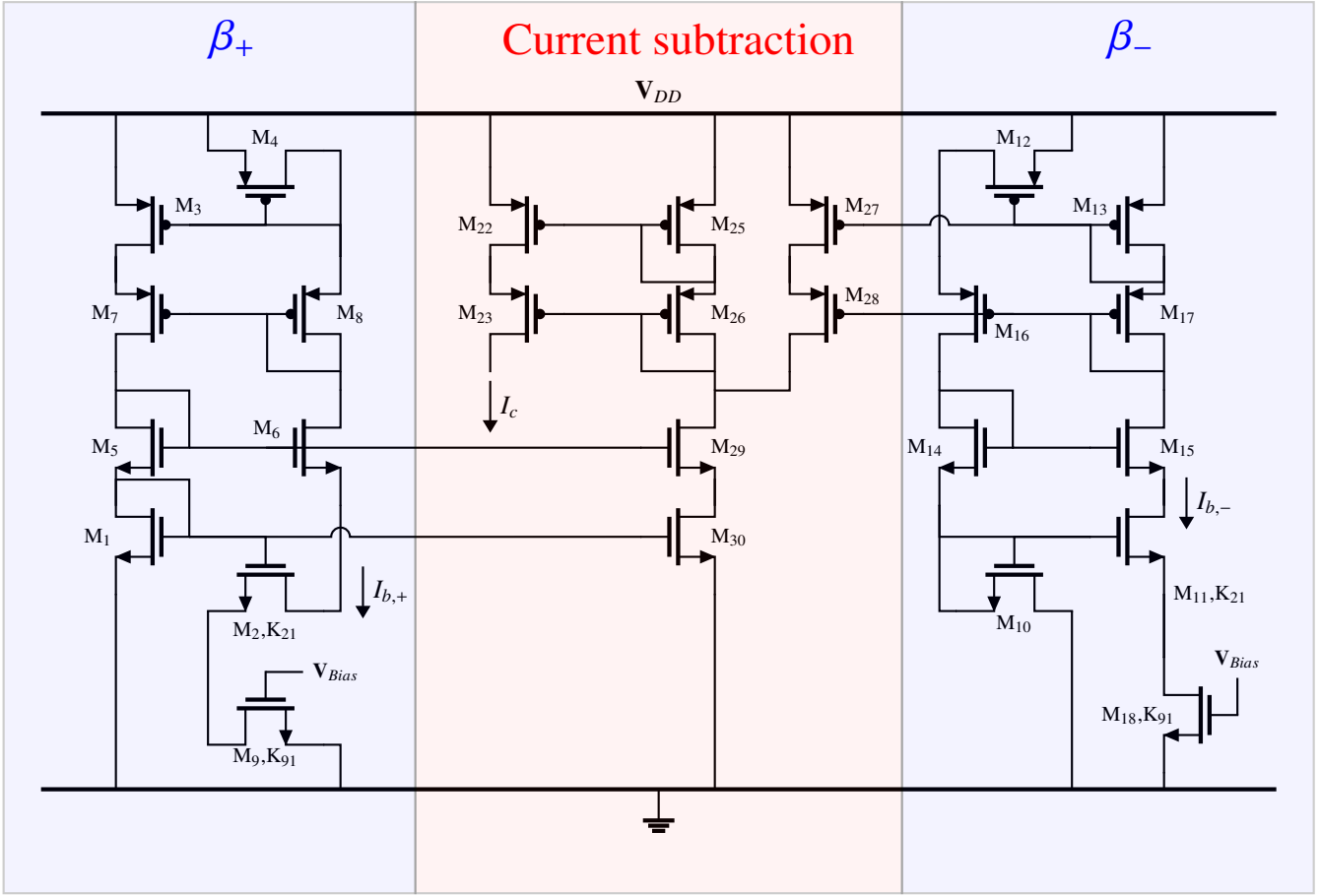

In this circuit, the two PMOS transistors M$_3$ and M$_4$ impose the same current in both in the absence of strain branches while the two NMOS transistors and the resistor impose a quadratic relation between the currents of the two branches. The transistors are assumed in saturation with long channel model and no mismatch between them. When the upper and lower parts of the circuits are connected, the equilibrium in the absence of strain leads to
\begin{equation}\EqFontSize
    I_a=2\frac{\left(1-1/\sqrt {K_{21}}\right)}{R^2\beta_{n,0}},
\end{equation}
where $R$ is the resistance, $K_{21}$ is the ratio between the width of the NMOS transistors, $\beta_{n,0}=\frac{W}{L}C_{ox}\mu_{n,0}$ with $C_{ox}$ the capacitance of the transistor gate, $W$ the width and $L$ the length of the transistor M$_1$. This last term leads to the name $\beta$-multiplier for the current reference.

Under strain conditions, the mobility of the different transistors is impacted. The expression of the reference current under uniaxial stress is then given by
\begin{equation}\EqFontSize\label{eq:Ia}
    I_a=2\frac{\left(e^{(\pi_3-\pi_1-\pi_4)\frac{\sigma}{2}}-e^{-\pi_2\frac{\sigma}{2}}/\sqrt {K_{21}}\right)~e^{-2\pi_r\sigma}}{R^2\beta_{n,0}},
\end{equation}
where $\pi_j$ is the piezoresistive coefficient of transistor M$_j$, $\pi_r$ is the piezoresistive coefficient of the resistor and $\sigma$ is the uniaxial stress applied on the circuit.

A second circuit is presented in Fig. \ref{fig:betamult}.b where the resistor is replaced with an active load. The transistor adds a control on the reference current by tuning the gate voltage $V_{bias}$. This tuning allows to control the power consumption of the circuit along with the sensitivities to strain and temperature. 

The reference current in this case can be expressed as
{\EqFontSize \begin{multline}\label{eq:Ib}
    I_b=2V_{ov}^2~\beta_{n,0} \cdot \\ \frac{\left( e^{(\pi_3-\pi_4-\pi_1)\frac{\sigma}{2}}-e^{-2\pi_2\sigma}/\sqrt K_{21} \right)^2}{\left[\left( e^{(\pi_3-\pi_4-\pi_1)\frac{\sigma}{2}}-e^{-2\pi_2\sigma}/\sqrt K_{21} \right)^2+e^{\pi_9\sigma}/\sqrt K_{91}\right]},
\end{multline}}
where $K_{91}$ is the ratio between the width of the transistors M$_9$ and M$_1$, and $V_{ov}=(V_{bias}-V_{th,n})$ with $V_{th,n}$ the threshold voltage of transistor M$_{9}$.

The complete development to find the expressions of the reference currents $I_a$ and $I_b$ can be found in Appendices \ref{app:A} and \ref{app:B}, respectively.

In both circuits, the PMOS and NMOS pairs (M$_1$/M$_2$, M$_3$/M$_4$ and M$_1$/M$_9$) are oriented perpendicularly to maximize the strain sensitivity. This configuration is necessary to observe the variations caused by the applied strain on the equilibrium current. 

\subsection{Current subtraction configuration}

It is possible to reach a larger strain sensitivity by combining circuits with positive and negative strain responses. Furthermore, a reduction of the temperature sensitivity can be achieved if the temperature responses of the two circuits are of the same sign thanks to the current subtraction approach \cite{Tang2003}. 
In order to obtain negative strain response, a rotation by 90° of the sensitive transistors in Fig \ref{fig:betamult}.b is made.
Two $\beta$-multiplier circuits can then be combined with a current subtraction circuit as displayed in Fig. \ref{fig:AddCircuit}. The sensing parts with positive and negative strain sensitivities are highlighted in blue while the current subtraction part is represented in red.

The output current of this circuit is given by
\begin{equation}\EqFontSize
    I_c=C_+~ I_{b,+}-C_-~ I_{b,-},
\end{equation}
where $I_{b,+}$ and $I_{b,-}$ are the current of the $\beta$-multiplier with positive and negative responses, respectively. $C_+$ and $C_-$ are constants that can be tuned by changing the ratio between the size of the MOSFETs of the $\beta$-multipliers and the current subtraction, i.e. M$_{27}$ and M$_{30}$. The linear combination of the two currents $I_{b,+}$ and $I_{b,-}$ with chosen constants ensures a positive output current $I_c$ regardless the applied strain.

\section{Experiment and discussion}\label{sec:experimental}
\subsection{Experimental set-up}
\begin{figure}[htb!]
    \centering
    \includegraphics[width=0.8\linewidth]{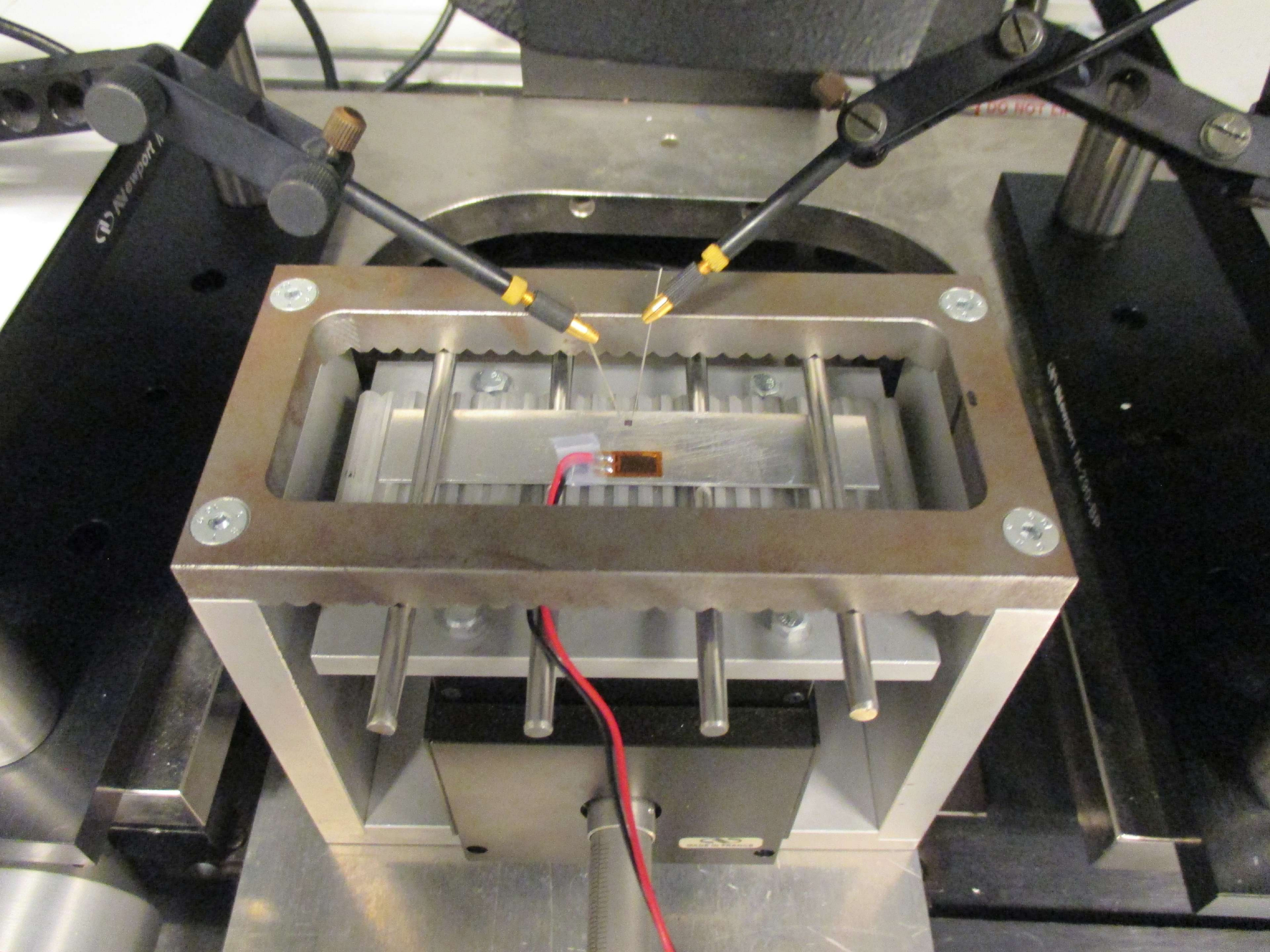}\vspace{0.2cm}
    \includegraphics[width=0.95\linewidth]{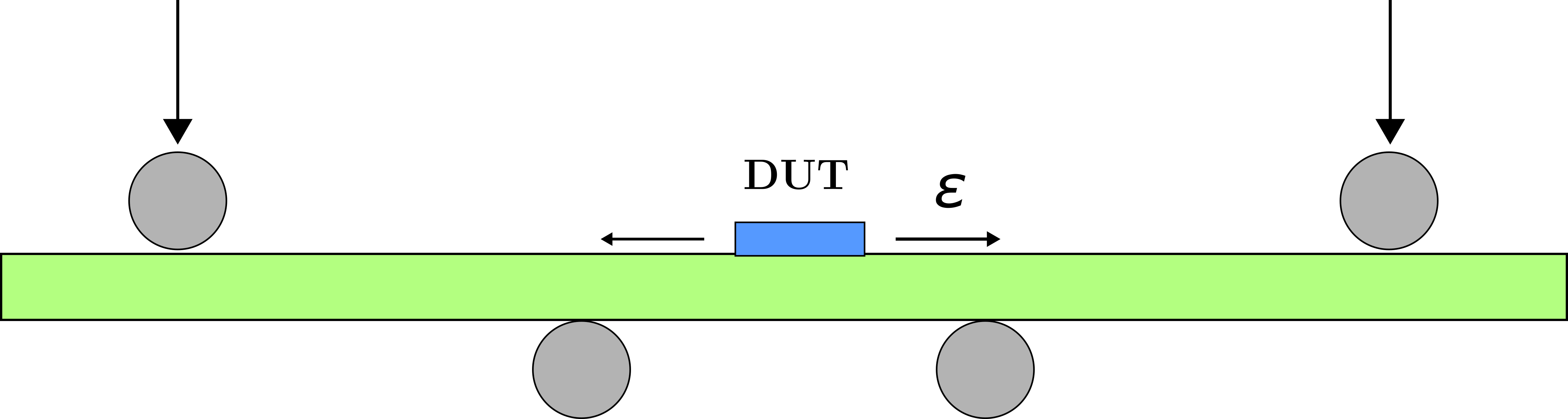}
    \caption{Illustration of the set-up for the four-point bending test.}
    \label{fig:FourPointBending}
\end{figure}

The circuits were fabricated using \textit{UMC L180} technology. The transistors have low threshold voltage with 1.8 V voltage maximum supply voltage. 
For the temperature measurements, a bare die was put on a heating stage while the strain measurements were performed in a four-point bending machine.
In this last case, the device under test (DUT) is ground down to a thickness of 50 um and next glued with \textit{M-Bond 200} adhesive from \textit{Vishay} on a 1 mm-thick aluminium strip. 
The strain is then applied using a four-point bending machine as displayed in Fig. \ref{fig:FourPointBending}. The bottom cylinders are spaced of 3 cm while a space of 8 cm is set for the the upper ones. The four-point bending method is a well-known mechanical test that allows a simple, stable and homogeneous deformation on a glued device \cite{Beaty1992}. A reference metallic strain gauge of 350 $\Omega$ from \textit{Micro-Measurements} is mounted next to the die in order to measure the strain applied to the device. The resistance of the gauge was measured with a Series 2000 digital multimeter from \textit{Keithley}.

The electrical measurements are made with a \textit{B-1500 Semiconductor Device Parameter Analyzer} from \textit{Keysight}. The electrical contact was made with tungsten probes directly on the die under test for both temperature or strain experiments.

The transistors are first measured separately in order to analyze the impact of strain and temperature on the mobility and the threshold voltage. Then, the currents of the four reference circuits are measured, i.e. the $\beta$-multiplier with resistor ($\beta_R$), the full-transistor $\beta$-multipliers with positive ($\beta_+$) and negative ($\beta_-$) strain responses and the current subtraction implementation ($\beta_{sub}$).

\subsection{Transistor analysis}
The tested NMOS and PMOS transistors have both a gate width of 8.5 $\mu$m and gate length of 5 $\mu$m. Each transistor type is duplicated and rotated by 90° in order to retrieve the transverse and longitudinal piezoresistive coefficients.

The significant parameters, i.e. the mobility and the threshold voltage, were extracted from I-V curves at different strain levels with the method from Jeppson \cite{Jeppson2013}. This method is based on a least-square fit that is stable and insensitive to the mobility degradation and series resistances. We measured the drain current and we varied the gate voltage between 0 and 1.8 V while the source and the drain were, respectively, put to ground (1.8 V) and 50 mV (1.75 V) the for the NMOS (PMOS) transistors (resp.). The body for both types of transistor is connected to the source.

The low-field mobilities obtained in the relaxed case at 25°C are 1131 cm$^2$V$^{-1}$s$^{-1}$ and 191 cm$^2$V$^{-1}$s$^{-1}$ for the NMOS and PMOS transistors, respectively. The threshold voltages in these conditions are, respectively, 73.6 mV and -330.5 mV  for the NMOS and PMOS transistor.

\begin{figure}[htb!]
    \centering
    \includegraphics[width=0.95\linewidth]{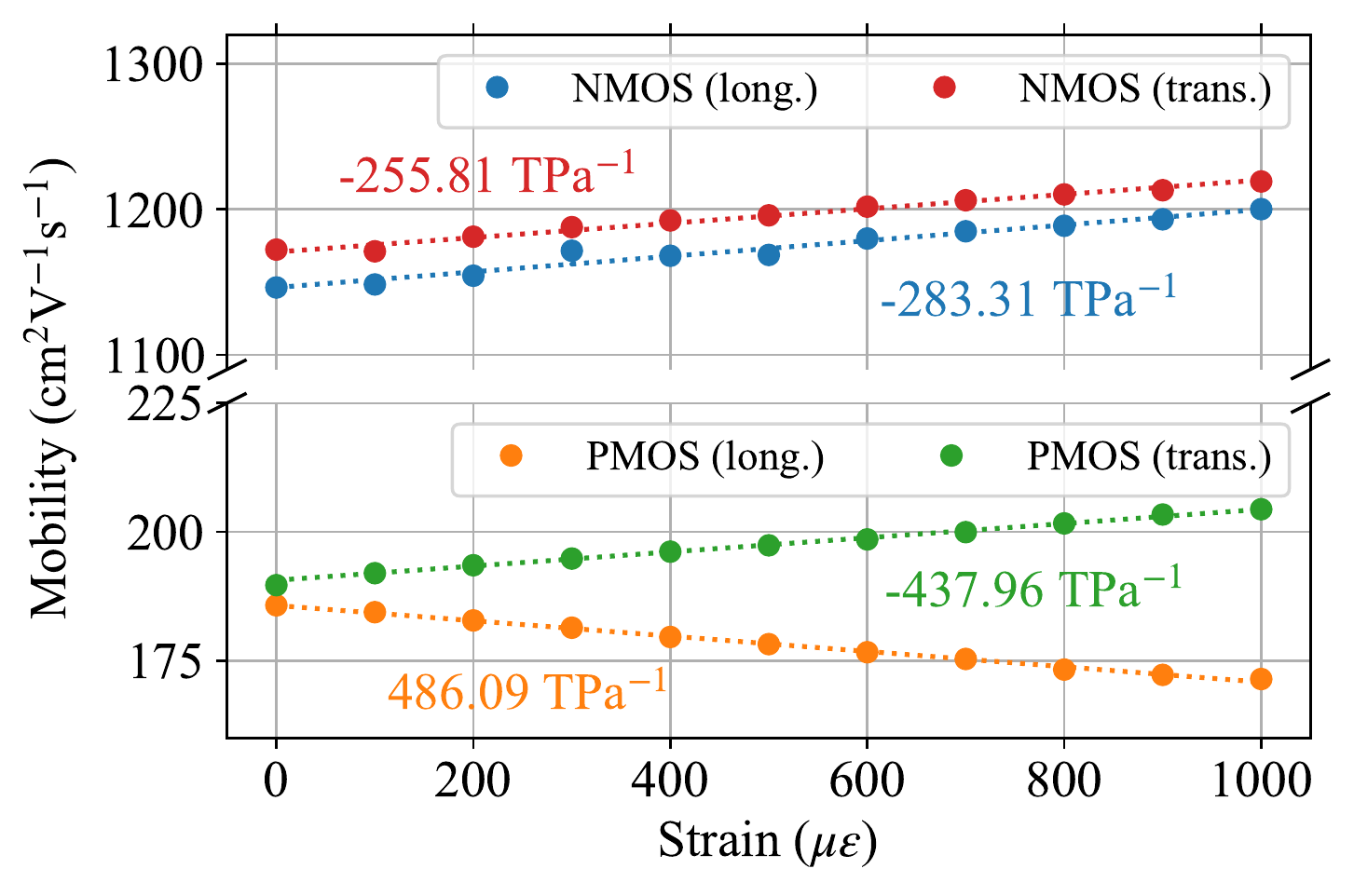}
    \caption{Mobility variation with regards to the applied strain for NMOS and PMOS transistors oriented in the longitudinal or transverse directions. The extracted piezoresistive coefficients are written next to the curves.}
    \label{fig:mustrain}
\end{figure}

The results of the strain measurements are displayed in Fig. \ref{fig:mustrain}. The piezoresistive coefficients are extracted by computing a linear interpolation on the mobility variation using relation \ref{eq:linear}. We obtain a transverse and longitudinal coefficients of -255 (-437) TPa$^{-1}$ and -283 (486) TPa$^{-1}$ for the NMOS (PMOS) transistors, respectively.
The PMOS transistor presents high and opposite piezoresistive coefficients while the coefficients for the NMOS transistor are smaller and of the same sign. 

Table \ref{tab:piezoresults} compares the experimentally measured data with piezoresistive coefficients published in the literature. The results  we obtained are in agreement with the ones shown in other works. The differences observed in the piezoresistive coefficients between the different works can be due to extrinsic perturbations that influence the strain response. Indeed, the bias condition as well as the regime of the measured transistors can have a strong impact on the extracted coefficients due to the different scattering mechanisms \cite{Jaeger2018}. 


\begin{table}[hbtp!]
    \renewcommand{\arraystretch}{1.3}
    \centering
    \begin{tabular}{c|ccccp{1.5cm}}\hline
         &$\pi_{n,t}$&$\pi_{n,l}$&$\pi_{p,t}$&$\pi_{p,l}$&Dimensions \\\hline
        This work&-255&-283&-437&486&WxL = 8.5 $\mu$m x 5 $\mu$m\\
       Wacker \textit{et al.} \cite{Wacker2011}&-470&-220&-450&520&WxL = 16 $\mu$m x 16 $\mu$m\\
       Bradley \textit{et al.} \cite{Bradley2001}&-250&-320&-385&415&L = 15 $\mu$m\\\hline
    \end{tabular}
    \vspace{0.2cm}
    \caption{Longitudinal and transverse piezoresistive coefficients in TPa$^{-1}$ experimentally measured and from the literature.}
    \label{tab:piezoresults}
\end{table}

\begin{figure}[ht!]
    \centering
    \includegraphics[width=0.95\linewidth]{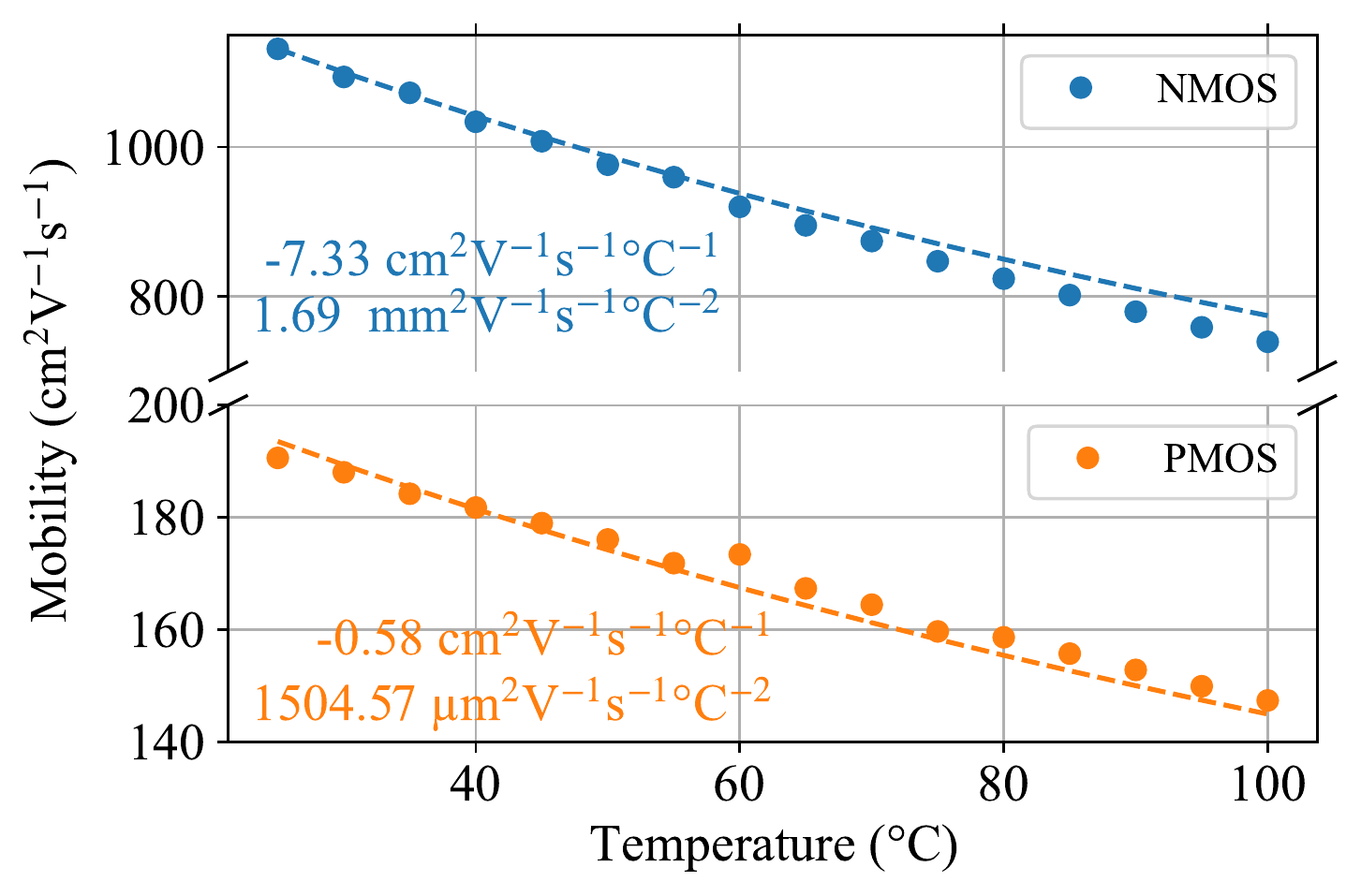}
    \caption{Mobility variation with regards to the temperature for NMOS and PMOS transistors. The temperature sensitivity extracted from the measurements is written next to it. The dashed lines represent the theoretical model from the transistor technology given by the UMC foundry.}
    \label{fig:mutemp}
\end{figure}

\begin{figure}[ht!]
    \centering
    \includegraphics[width=0.95\linewidth]{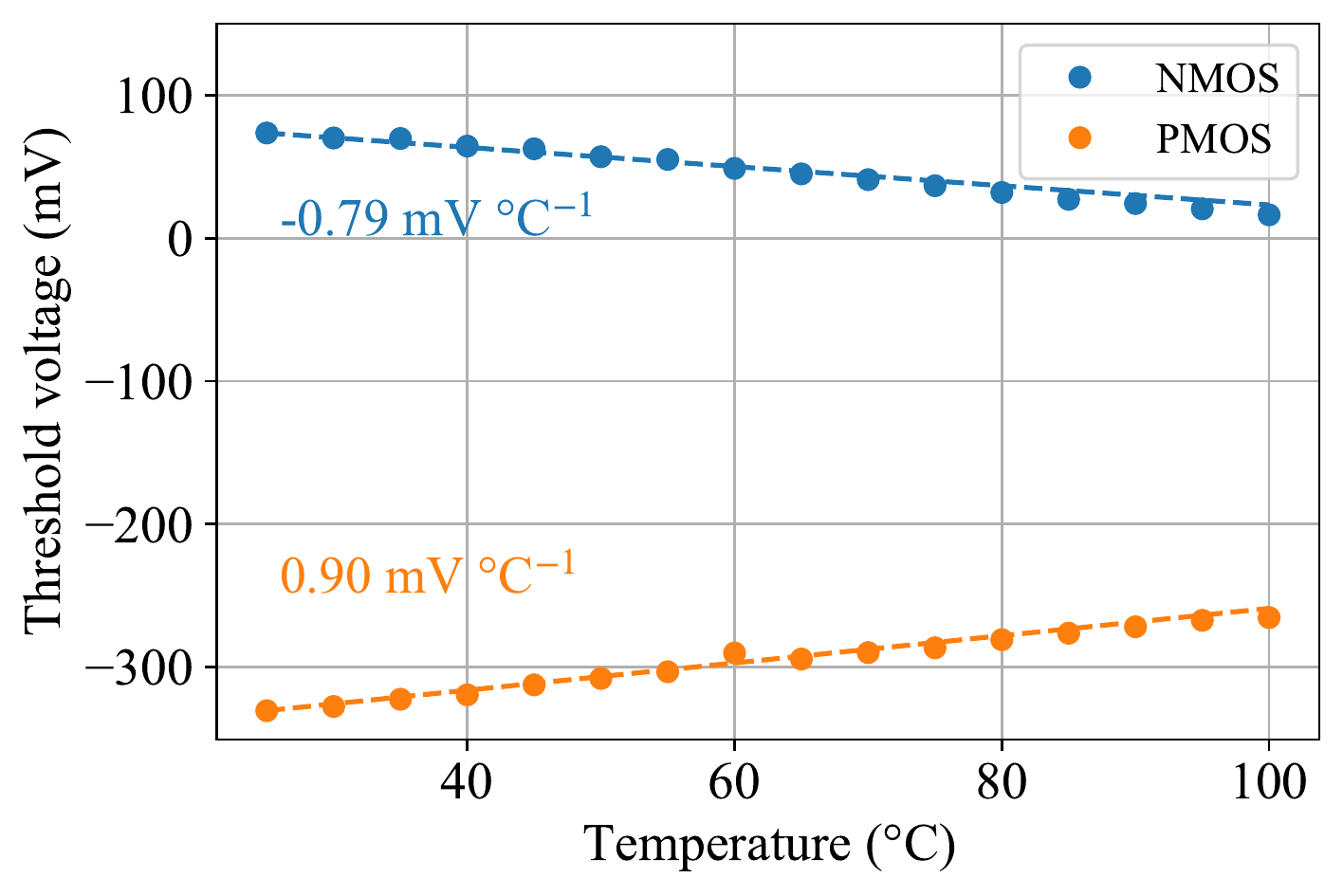}
    \caption{Threshold voltage with regards to the temperature for NMOS and PMOS transistors. The temperature sensitivity extracted from the measurements is written next to it. The dashed lines represent the theoretical model from the transistor technology given by the UMC foundry.}
    \label{fig:vthtemp}
\end{figure}

The results of the temperature measurements are displayed in Fig. \ref{fig:mutemp} and \ref{fig:vthtemp}. Mobility and threshold voltage sensitivities of -7.33 cm$^2$V$^{-1}$s$^{-1}$°C$^{-1}$/1.69 mm$^2$V$^{-1}$s$^{-1}$°C$^{-2}$ and -0.79 mV°C$^{-1}$ are, respectively, found for n-type transistors while the p-type transistors show a mobility sensitivity of -0.58 cm$^2$V$^{-1}$s$^{-1}$°C$^{-1}$/1504.57 \textmu m$^2$V$^{-1}$s$^{-1}$°C$^{-2}$ and a threshold voltage sensitivity of 0.90 mV°C$^{-1}$. The dashed curves are computed using the theoretical model provided by the circuit supplier. The curves are in good agreement with the experimental work for both NMOS and PMOS transistors.


\subsection{$\beta$-multiplier analysis}
The $\beta$-multiplier circuits are tested similarly with the four-point bending method. 
Strain and temperature measurements are conducted on four circuits where the output current is measured according to the supply voltage for the implementation with resistor or bias voltage for the full-transistor ones. The first circuit is the $\beta$-multiplier with resistor ($\beta_R$) displayed in Fig. \ref{fig:betamult}.a where the output current is $I_a$. The resistor is made of poly-silicon and reaches a value of 10 k$\Omega$. The configuration with an active load is then investigated with two transistor orientations leading to positive ($\beta_+$) and negative ($\beta_-$) strain sensitivities. The output currents of these circuits represented in Fig. \ref{fig:betamult}.b are $I_{b,+}$ ($I_{b,-}$) for the positive (negative) contribution. Finally, the output current $I_c$ of the circuit with current subtraction ($\beta_{sub}$) represented in Fig. \ref{fig:AddCircuit} is measured.

The gauge factor $GF$ is used to analyze and compare the strain sensitivity. This factor is defined as the relative current variation with regards to the strain, i.e.
\begin{equation}\EqFontSize
    \frac{\Delta I}{I(\varepsilon=0)}=GF\cdot\varepsilon
\end{equation}

\begin{figure}[h!]
    \centering
    \includegraphics[width=0.95\linewidth]{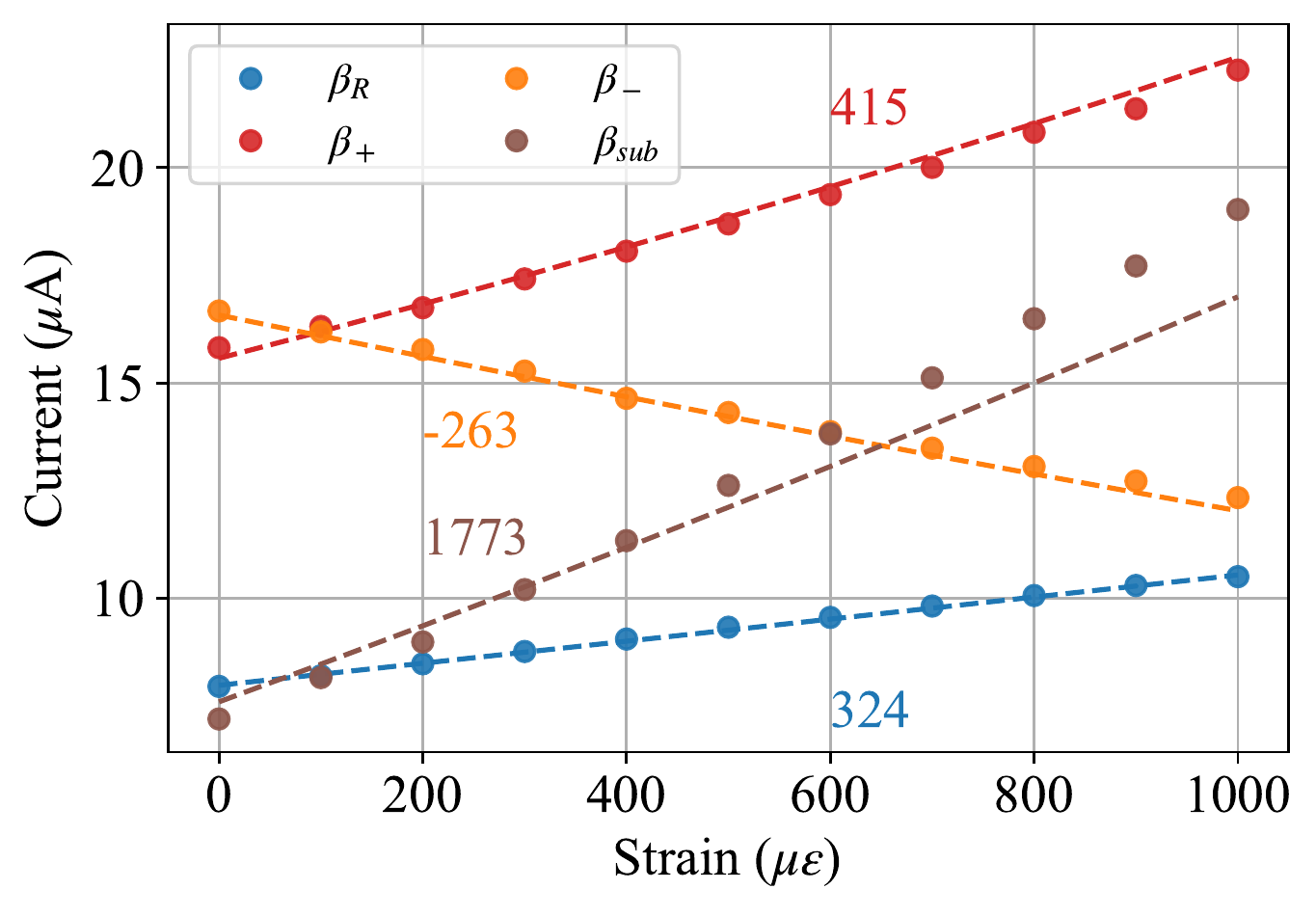}
    \caption{Output current with regards to the strain of the different implementations at $V_{DD}=1.8$ V and $V_{bias}=$ 1.2 V, i.e. the $\beta$-multiplier with resistor ($\beta_R$), the full-transistor $\beta$-multipliers with positive ($\beta_+$) and negative ($\beta_-$) strain responses and the current subtraction implementation ($\beta_{sub}$). The gauge factors extracted from the measurements are written next to the curves. The dashed lines represent the theoretical results obtained with the piezoresistive coefficients inserted in the analytical relations.}
    \label{fig:Ifinal}
\end{figure}

\begin{figure}[h!]
    \centering
    \includegraphics[width=0.95\linewidth]{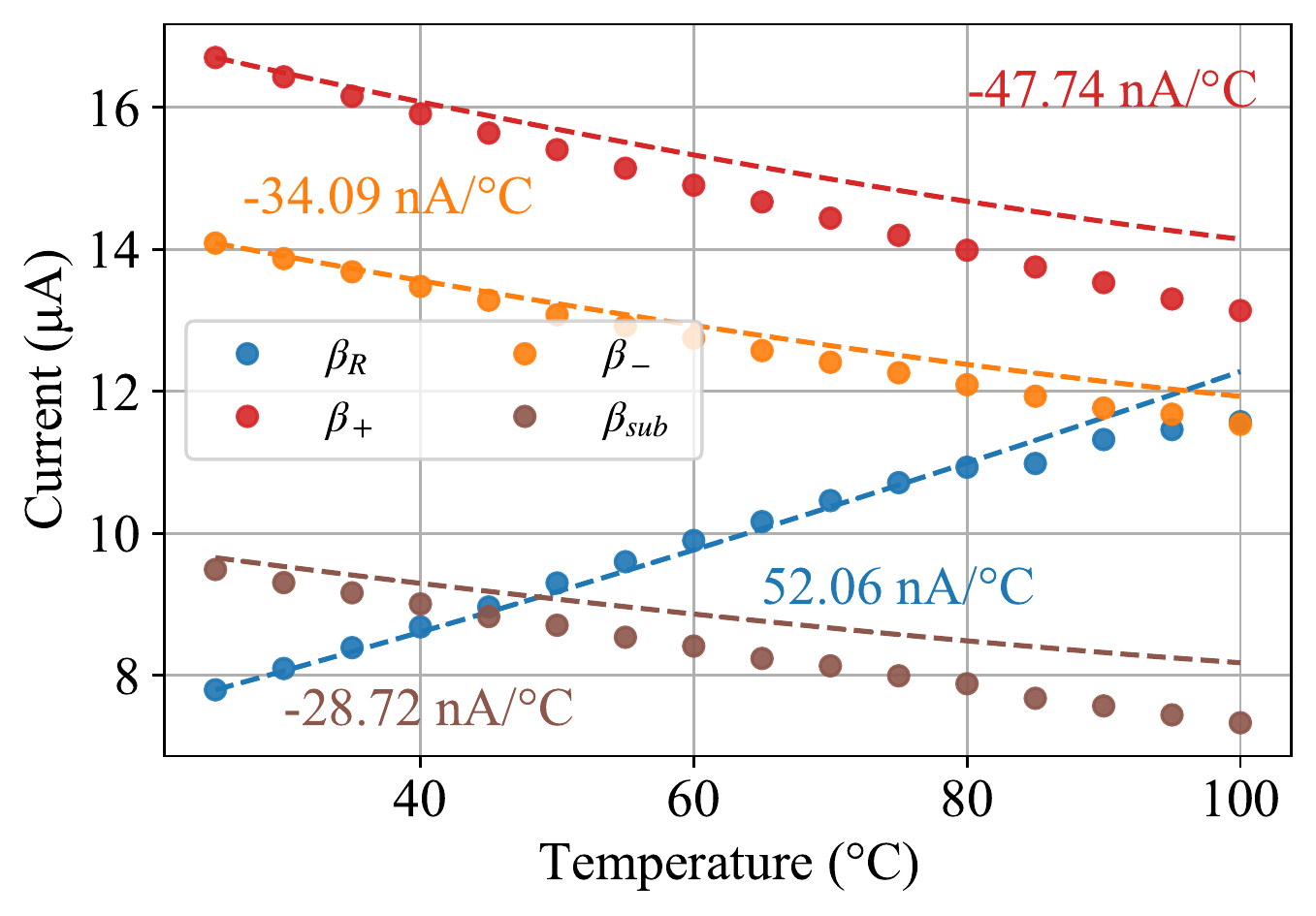}
    \caption{Output current with regards to the temperature of the different implementations at $V_{DD}=1.8$ V and $V_{bias}=$ 1.2 V, i.e. the $\beta$-multiplier with resistor ($\beta_R$), the full-transistor $\beta$-multipliers with positive ($\beta_+$) and negative ($\beta_-$) strain responses and the current subtraction implementation ($\beta_{sub}$). The temperature sensitivities extracted from the measurements are written next to the curves. The dashed lines represent the theoretical results obtained by simulation using the spice model of the UMC180 transistors given by the UMC foundry.}
    \label{fig:Itemp}
\end{figure}

The current variations under strain stimuli are displayed in Fig. \ref{fig:Ifinal} while the variations according to the temperature are shown in Fig. \ref{fig:Itemp}. 

The dashed curves represent the theoretical results obtained by simulations for the temperature data and analytically for the strain results. The simulations are made using the spice model given by the UMC foundry for the UMC180 transistors while the strain impact is computed by injecting the piezoresistive coefficients found (cfr. Table \ref{tab:piezoresults}) in relations (\ref{eq:Ia}) and (\ref{eq:Ib}).

\subsubsection{$\beta-$multiplier with resistor (Fig. \ref{fig:betamult}.a)}

\begin{figure}[h!]
    \centering
    \includegraphics[width=1\linewidth]{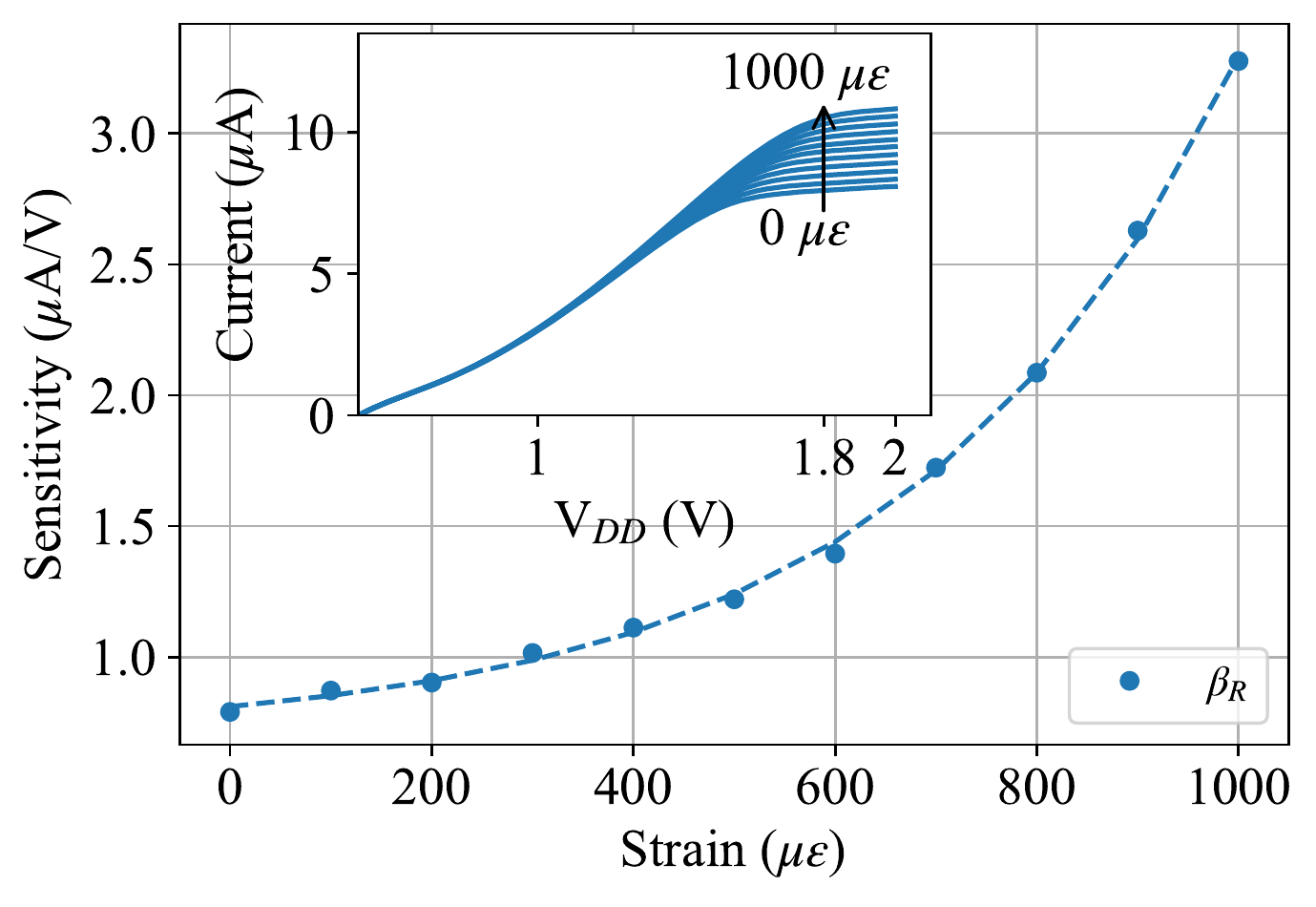}
    \caption{Output current sensitivity to supply voltage of the $\beta$-multiplier with resistor ($\beta_R$) with regards to the applied strain. The inset represents the raw results with the arrow at the supply voltage where the sensitivity is computed (1.8 V).}
    \label{fig:IRvdd}
\end{figure}

\begin{figure}[h!]
    \centering
    \includegraphics[width=1\linewidth]{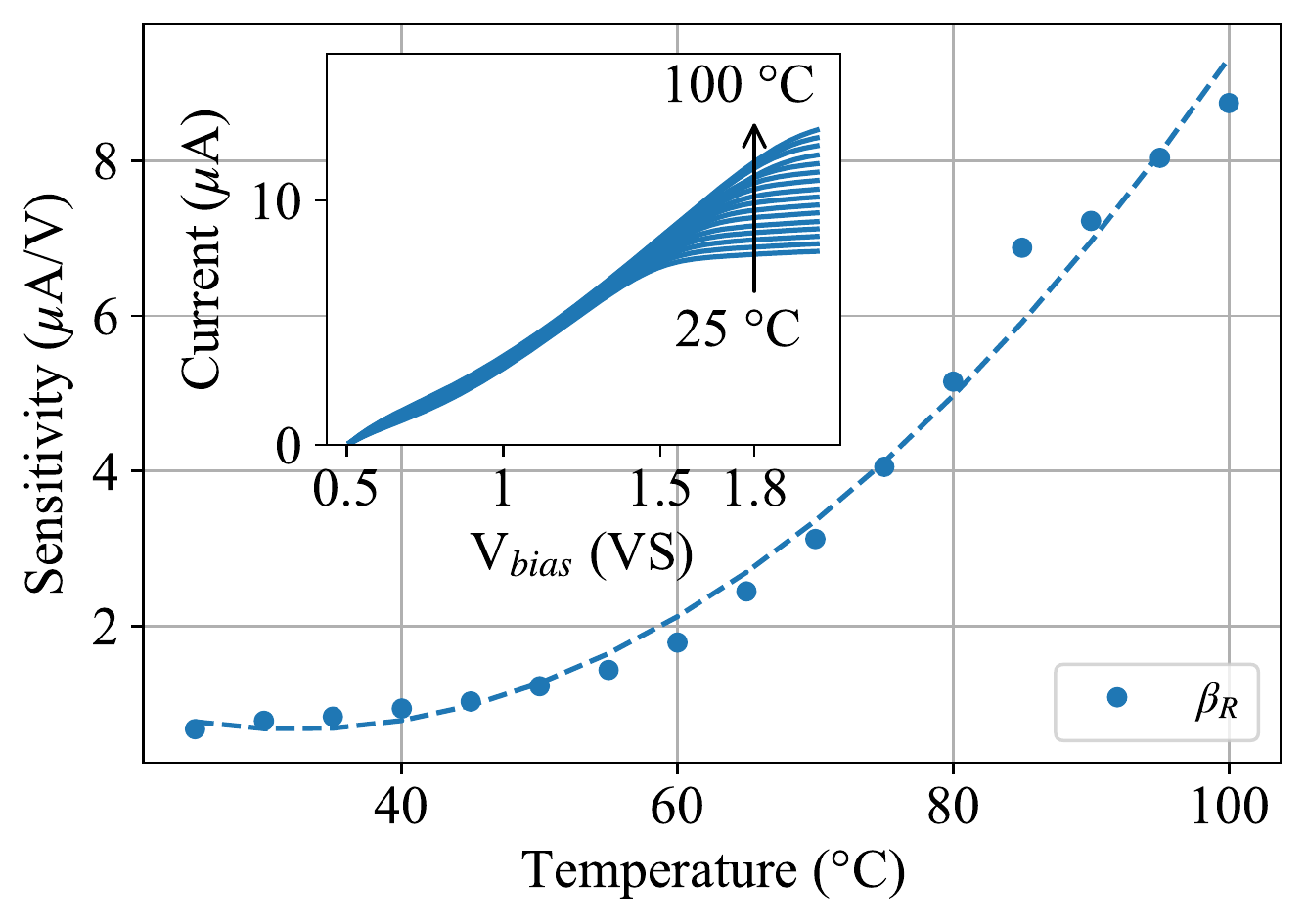}
    \caption{Output current sensitivity to supply voltage of the $\beta$-multiplier with resistor ($\beta_R$) with regards to the temperature. The inset represents the raw results with the arrow at the supply voltage where the sensitivity is computed (1.8 V).}
    \label{fig:IRVdd_temp}
\end{figure}

The implementation $\beta_R$ shows a gauge factor of 324 (2.59 nA/$\mu\varepsilon$) and temperature sensitivity of 52.06 nA/°C with a supply voltage of 1.8 V. The sensitivity to supply voltage of the implementation is investigated in Fig. \ref{fig:IRvdd}. A sensitivity of 0.79 $\mu$A/V is obtained in the relaxed case. The sensitivity increased with the strain due to the voltage limit to keep the transistors in saturation regime being closer to the operating voltage of 1.8 V. Under high strain condition of 1000 $\mu\varepsilon$, the sensitivity is up to 3.28  $\mu$A/V.
The effect of the temperature on the supply sensitivity presents the same behavior and is displayed in Fig. \ref{fig:IRVdd_temp}. We obtained a sensitivity of 0.79 $\mu$A/V at 25°C that increases up to 8.74 $\mu$A/V at 100°C. 
The cascode implementation allows low supply sensitivity but brings the bias limit of the circuit close to the operating point.

\subsubsection{$\beta-$multiplier with active load (Fig. \ref{fig:betamult}.b) and subtraction circuit (Fig. \ref{fig:AddCircuit})}

\begin{figure}[htb!]
    \centering
    \includegraphics[width=0.95\linewidth]{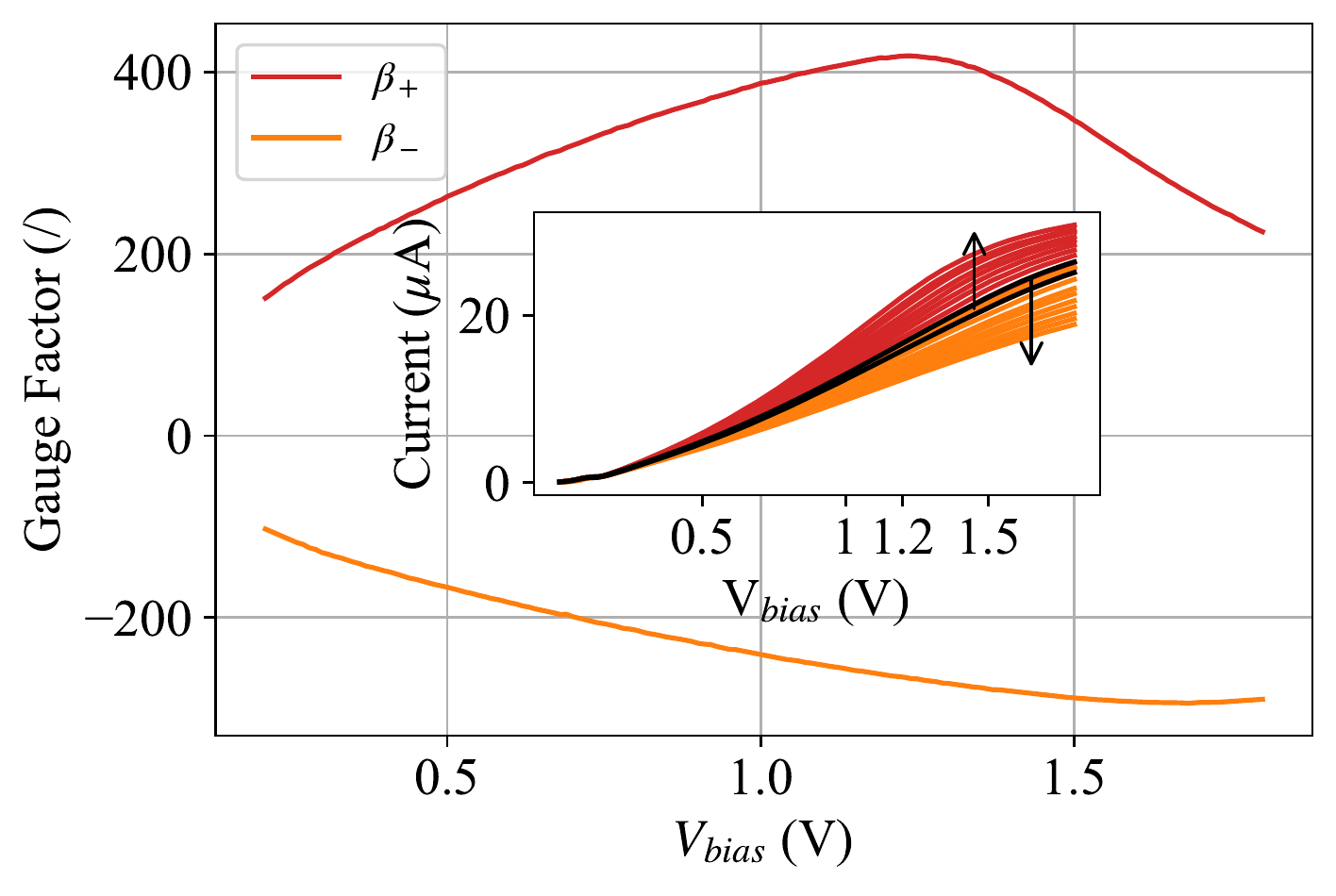}
    \caption{Gauge factor computed according to the bias voltage for the full-transistor implementations in strong inversion with positive ($\beta_+$) and negative ($\beta_-$) responses. The inset presents the raw I-V results with the arrow indicating the direction of the strain increase from 0 to 1000 $\mu\epsilon$.}
    \label{fig:Ivbias}
\end{figure}

\begin{figure}[h!]
    \centering
    \includegraphics[width=0.95\linewidth]{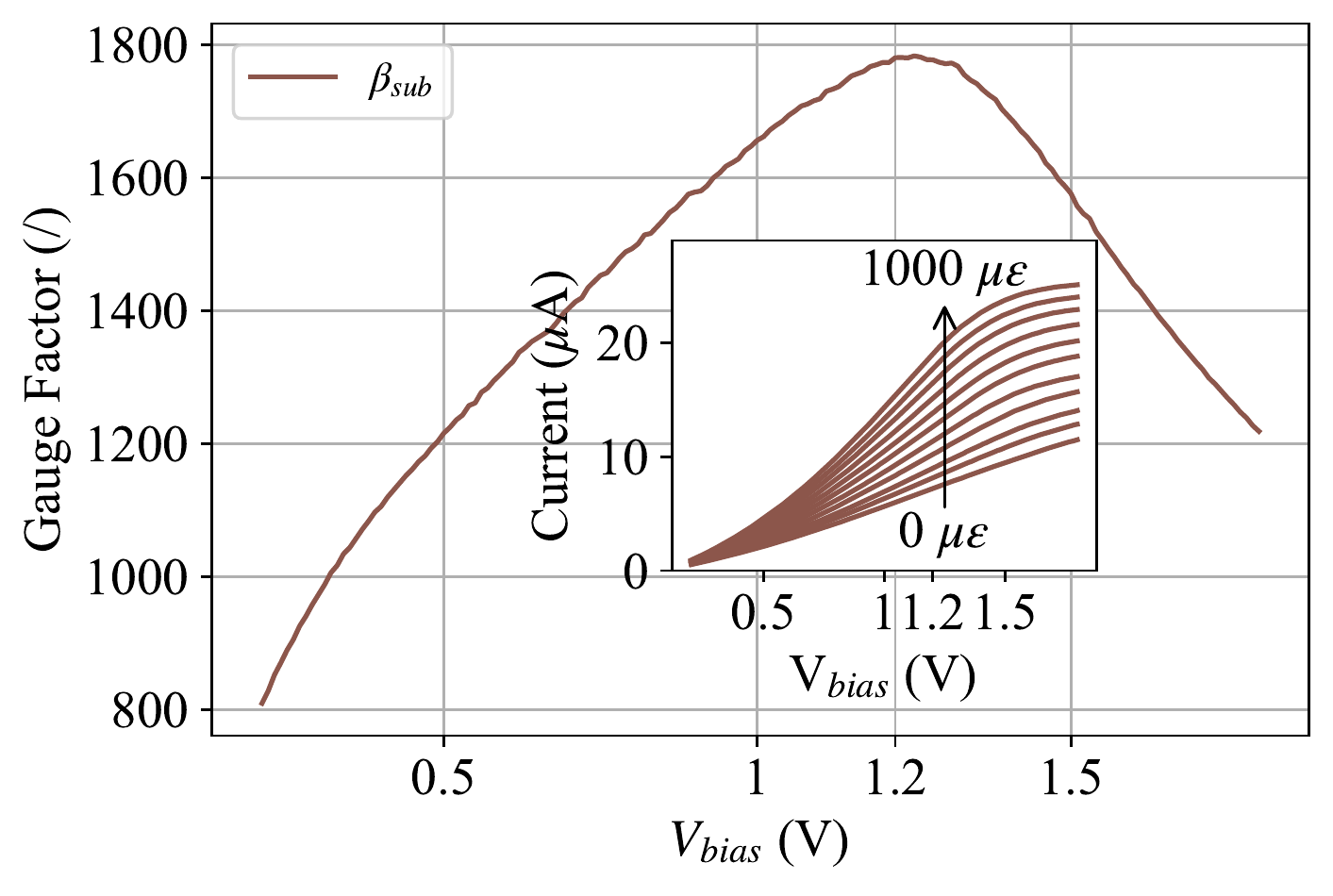}
    \caption{Gauge factor computed according to the bias voltage for the full-transistor current subtraction implementation ($\beta_{sub}$) in strong inversion. The inset presents the raw I-V results with the arrow indicating the direction of the strain increase from 0 to 1000 $\mu\epsilon$.}
    \label{fig:Ioutvbias}
\end{figure}

The implementation $\beta_-$ and $\beta_+$ show a gauge factor of -263 (-4.37 nA/$\mu\varepsilon$) and 415 (6.47 nA/$\mu\varepsilon$) depending on the transistor orientation while the temperature sensitivities are -34.09 nA/°C and -47.74 nA/°C, respectively. 

The active load allows a current control by tuning the bias voltage. As consequence, the strain and temperature sensitivities of the output current depend on the voltage applied. The gauge factor variations according to the bias voltage is shown in Fig. \ref{fig:Ivbias}. A maximum gauge factor of 415 (resp. -290) around 1.2 V (resp. 1.6 V) is found for $\beta_+$ (resp. $\beta_-$) implementation. At higher bias voltage, the current flowing through the devices increases the gate voltage of the current mirrors. This bring the transistors into the triode regime and degraded the gauge factor of the circuit.

\begin{figure}[h!]
    \centering
    \includegraphics[width=0.95\linewidth]{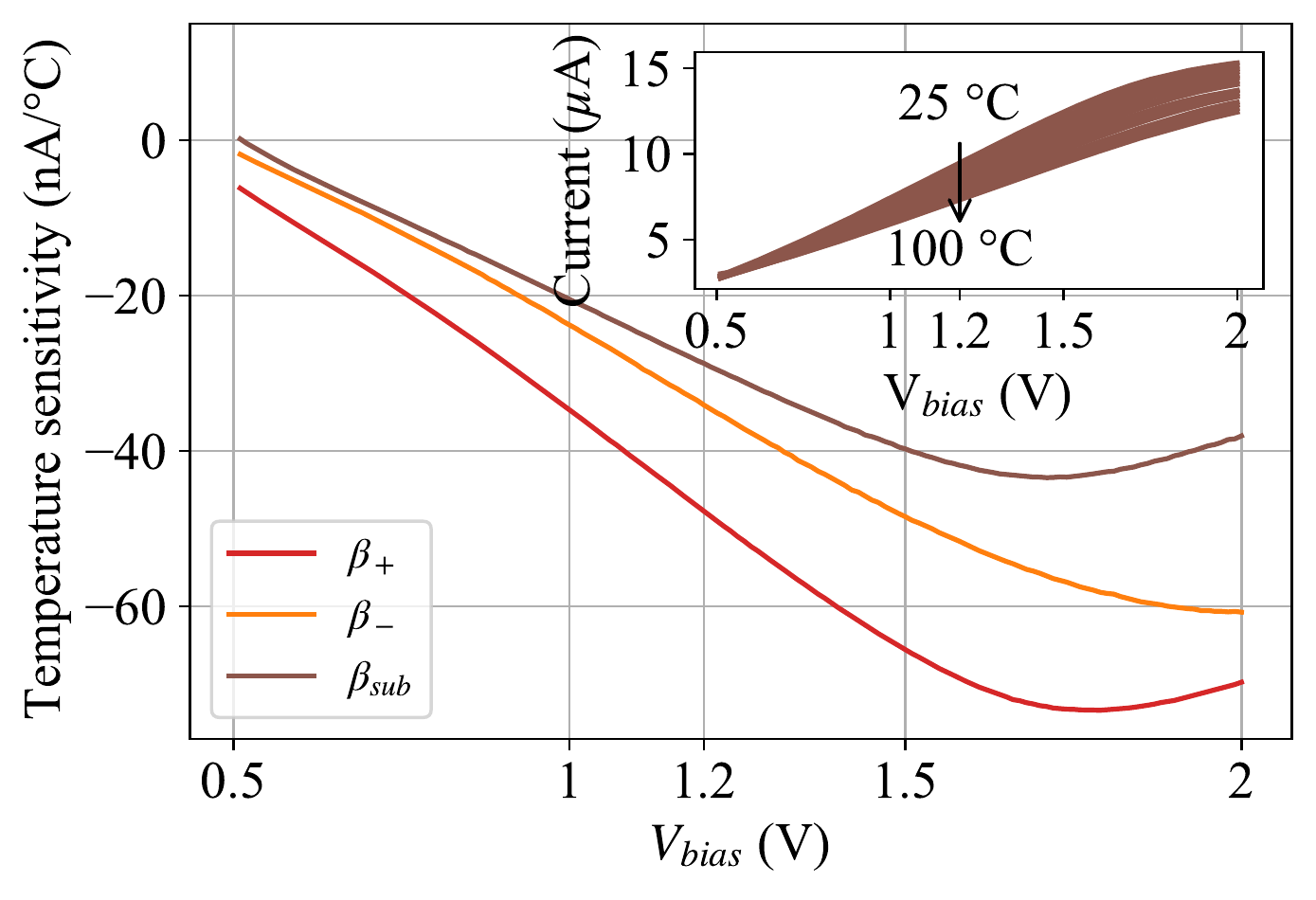}
    \caption{Temperature sensitivity according to the bias voltage for full-transistor implementations in strong inversion, i.e. $\beta$-multiplier with positive ($\beta_+$) and negative ($\beta_-$) strain response combined with a current subtraction circuit ($\beta_{sub}$). The inset shows the raw I-V curves of the measurements for the current of $\beta_{sub}$.}
    \label{fig:dIdTemp}
\end{figure}

The temperature sensitivity depending on the bias voltage is represented in Fig. \ref{fig:dIdTemp}. At 1.2 V, sensitivities of -47.74 nA/°C and -34.09 nA/°C are obtained for the output currents of $\beta_+$ and $\beta_-$, respectively. 


The circuit with current subtraction $\beta_{sub}$ presents the highest factor with 1773 (12.02 nA/$\mu\varepsilon$) and a temperature sensitivity of -28.72 nA/°C for a bias voltage of 1.2 V.
The gauge factor of the current subtraction circuit is displayed in Fig. \ref{fig:Ioutvbias}.

As the temperature sensitivities are both negative for $\beta_+$ and $\beta_-$ circuits, the current subtraction leads to lower temperature sensitivity. On top of that, we obtained strain sensitivity nearly three times higher.

\begin{figure}[h!]
    \centering
    \includegraphics[width=0.95\linewidth]{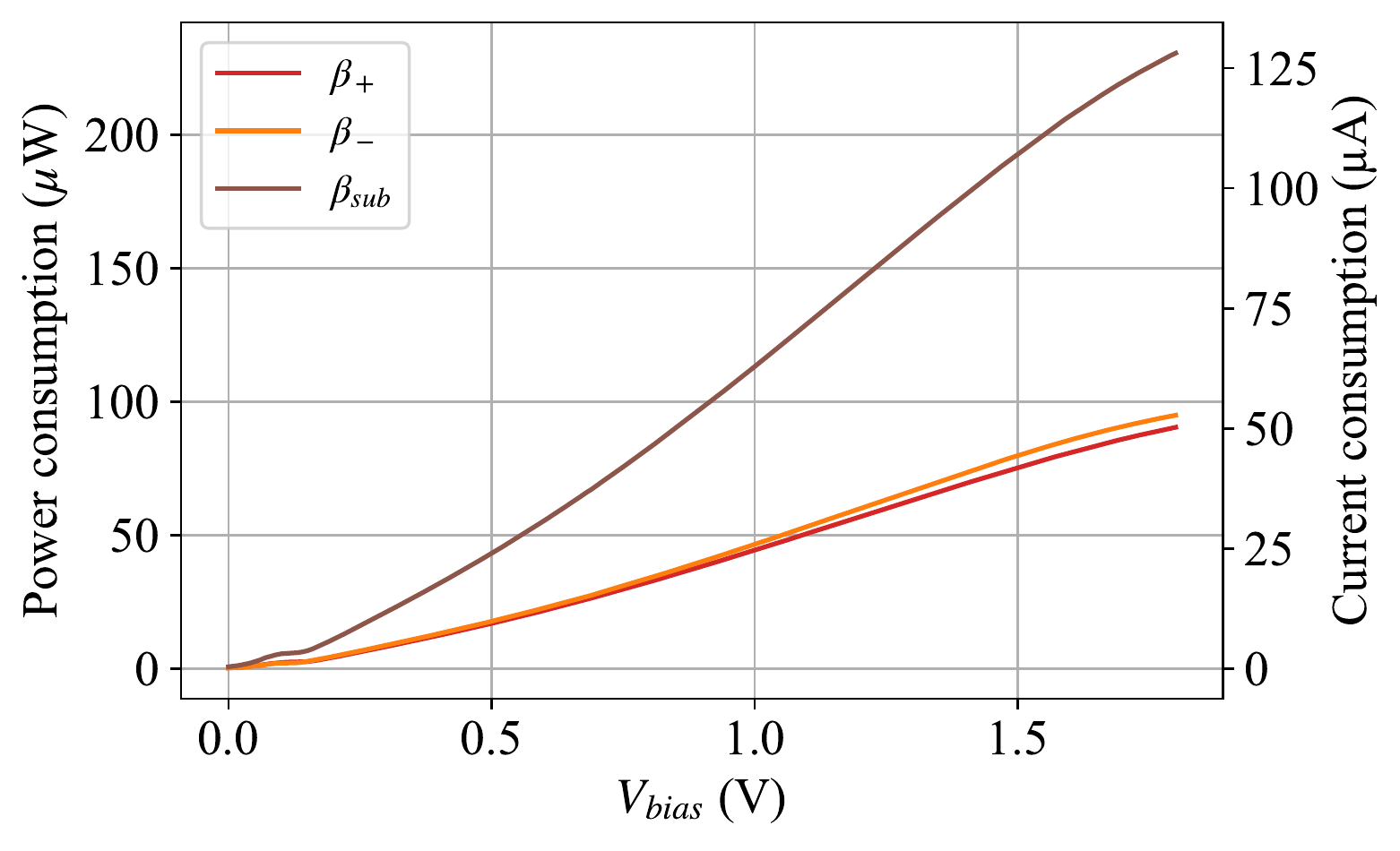}
    \caption{Power and current consumption according to the bias voltage for full-transistor implementations, i.e. $\beta$-multiplier with positive ($\beta_+$) and negative ($\beta_-$) strain response combined with a current subtraction circuit ($\beta_{sub}$).}
    \label{fig:PVbias}
\end{figure}
\begin{table*}[h!]
    \centering
    \renewcommand{\arraystretch}{1.3}
    \begin{tabular}{c|cccc}\hline
        Implementation &$\beta_R$& $\beta_+$ & $\beta_-$ & $\beta_{sub}$ \\\hline
        Strain sensitivity (nA/$\mu\varepsilon$)&2.59&6.47&-4.37&12.02\\
        Gauge Factor (/)&324&415&-263&1773\\
        Temperature sensitivity (nA/°C)&52.06&-47.74&-34.09&-28.72\\
        Power Consumption ($\mu$W)&28.6&56.96&60.02&145.45\\\hline
    \end{tabular}
    \vspace{0.2cm}
    \caption{Summary of the main performances at $V_{DD}=1.8$ V and $V_{bias}=$ 1.2 V of the different implementations, i.e. the $\beta$-multiplier with resistor ($\beta_R$), the full-transistor $\beta$-multipliers with positive ($\beta_+$) and negative ($\beta_-$) strain responses and the current subtraction implementation ($\beta_{sub}$).}
    \label{tab:performances}
\end{table*}

The improvement of the performances is however counterbalanced by the larger power consumption of 145.45 $\mu$W. It is possible to reduce the current consumption (and therefore the power consumed) of the circuit by reducing the bias voltage as displayed in Fig. \ref{fig:PVbias}. Indeed, a consumption of around 50 $\mu$W can for example be obtained with a bias voltage of 0.5 V but with a gauge factor of around 1200 instead of 1773.

Table \ref{tab:performances} summarizes the performances of the different implementations. The modifications of the $\beta_R$ implementation allow for reaching higher strain sensitivity and gauge factor (from 324 to 1773) at the cost of higher power consumption (from 28.6 $\mu$W to 145 $\mu$W). Furthermore, the more advanced solution $\beta_{sub}$ also shows better temperature sensitivity compared with the basic reference circuit (from 52.06 nA/°C to -28.72 nA/°C).

\section{Conclusion}
In this work, we developed a new strain sensor based on $\beta$-multiplier topology in order to reduce the device sensitivity to supply voltage. The piezoresistive effect in silicon is exploited directly in the elements composing the circuit. This allows the circuit to work directly as the sensing element, leading to high performances and easy-to-measure output.
We presented a reference circuit adapted for strain sensing applications by changing the relative orientations of the transistors. 
The strain impact on the reference currents is analytically described and verified with the experimental results. The theoretical methodology developed shows the possibility to easily compute the strain impact in reference circuit. The analytical work should allow further predictions of the performances of the circuit depending on the elements and device orientations.

The transistors used in the circuits were first characterized to predict the response of the different implementations. The basic topology with resistor gives a strain sensitivity of 2.54 nA/$\mu\epsilon$ (gauge factor of 324) with a low power consumption of 28.6 $\mu$W. The temperature sensitivity was 52.06 nA/°C. This solution was improved with full-transistor implementation combined in a current subtraction circuit. This leads to high strain sensitivity of 12.02 nA/$\mu\epsilon$ (gauge factor of 1773) but at the cost of higher power consumption of 145.45 $\mu$W. The temperature sensitivity was lower down to -28.72 nA/°C with the subtraction principle. Furthermore, the full-transistor principle allows to tune the performances of the circuit with a gauge factor between 800 and 1800 for a power consumption between 20 and 200 $\mu$W.

%

\appendix
\section{Output current in $\beta$-multiplier with resistor}\label{app:A}
By applying Kirchhoff's voltage law on the resistor and NMOS transistors M$_1$ and M$_2$ from Fig. \ref{fig:betamult}.a, the following relation is obtained 

\begin{equation}\EqFontSize\label{beta1}
V_{GS,1}=V_{GS,2}+R I_{D,2}.
\end{equation}
where $V_{GS,i}$ and $I_{D,i}$ stand for the gate-to-source voltage and drain current of transistor M$_i$, respectively.
If the transistors are working in saturation and by neglecting the Early effect,  $V_{GS,i}$ is given by 

\begin{equation}\EqFontSize\label{beta2}
V_{GS,i}=V^{th}_{n,i}+\sqrt{\frac{2I_{D,i}}{\beta_{n,i}(\sigma)}},
\end{equation}
where $V^{th}_{n,i}$ is the threshold voltage of transistor $i$ and $\beta_{j,i}(\sigma)=\left(\frac{W}{L}\right)_i \mu_{n,i} C_{ox}$ with $j$ standing for the electrons (n) or holes (p) contribution. 
At the level of the PMOS transistors M$_3$ and M$_4$, the current mirror configuration gives 

\begin{equation}\EqFontSize\label{beta3}
\left\{ \begin{array}{l}
V_{SG,3}=V_{SG,4} \\
I_{D,2}=I_{D,4} \\
I_{D,3}=I_{D,1}
\end{array} \right. \Rightarrow \frac{I_{D,1}}{\beta_{p,3}(\sigma)}=\frac{I_{D,2}}{\beta_{p,4}(\sigma)}.
\end{equation}

The currents of the transistors M$_3$ and M$_1$ are equal, the same goes for transistors M$_2$ and M$_4$.

Substituting relations (\ref{beta2}) and (\ref{beta3}) in relation (\ref{beta1}), we obtain 
\begin{equation}\EqFontSize
V^{th}_{n,2}+\sqrt{\frac{2I_{D,2}}{\beta_{n,2}(\sigma)}}+R I_{D,2} =V^{th}_{n,1}+\sqrt{\frac{2I_{D,2}}{\beta_{n,1}(\sigma)} \frac{\beta_{p,3}(\sigma)}{\beta_{p,4}(\sigma)}},
\end{equation}
\begin{multline*}\label{eq:mismatch2}
R I_{D,2}-\sqrt{I_{D,2}}\underbrace{\sqrt{2}\left(\sqrt{\frac{\beta_{p,3}(\sigma)}{\beta_{p,4}(\sigma)\beta_{n,1}(\sigma)}}-\frac{1}{\sqrt{\beta_{n,2}(\sigma)}} \right)}_{f(\sigma)}\\+\Delta V^{th}_n=0,
\end{multline*}

\begin{equation}\EqFontSize
\sqrt{I_{D,2}}=\frac{f(\sigma)\pm\sqrt{f(\sigma)^2+4R\Delta V^{th}_n}}{2R}.
\end{equation}

By assuming no threshold voltage mismatch between the transistors, i.e. $\Delta V^{th}_n=0$, we obtain
\begin{equation}\EqFontSize \label{beta4}
I_{D,2}= \left\{ \begin{array}{l}
0\\
\frac{f(\sigma)^2}{R^2}
\end{array} 
\right. .
\end{equation}

Equation (\ref{beta4}) shows two solutions for the current $I_{D,2}$. In order to avoid the zero-current solution, a start-up circuit is thus needed. 

Using relation (\ref{eq:muexp}), the current is given by
\begin{equation}\EqFontSize
I_{D,2} = \frac{2\left(\sqrt{\frac{\beta_{p,3}(\sigma)}{\beta_{p,4}(\sigma)\beta_{n,1}(\sigma)}}-\frac{1}{\sqrt{K_{21}\beta_{n,2}(\sigma)}}\right)^2}{R^2}.
\end{equation}

We used the exponential relation for the gain, i.e. $\beta(\sigma)=\beta^0\cdot e^{-\pi\sigma}$, to find
\begin{equation}\EqFontSize
    I_{D,2}=2\frac{\left(e^{-(\pi_3-\pi_1-\pi_4)\frac{\sigma}{2}}-e^{\pi_2\frac{\sigma}{2}}/\sqrt {K_{21}}\right)~e^{-2\pi_r\sigma}}{R^2\beta^{0}},
\end{equation}
with $\beta^0=\beta_{n,1}(\sigma=0)=\beta_{p,3}(\sigma=0)=\beta_{p,4}(\sigma=0)=\frac{\beta_{n,2}(\sigma=0)}{K_{21}}$. $\pi_i$  is the piezoresistive coefficient of transistor M$_i$ and $\pi_r$ is the one of the resistor.

In the absence of strain, we find the classical expression for the reference current of the $\beta$-multiplier:
\begin{equation}\EqFontSize
    I_{D,2}=2\frac{\left(1-1/\sqrt {K_{21}}\right)}{R^2\beta^{0}}.
\end{equation}

\section{Output current in full-transistor $\beta$-multiplier}\label{app:B}
A similar development than in Appendix \ref{app:A} can be done for the full-transistor circuit. In this case, the resistor is replaced by a transistor (M$_9$) in triode region as displayed in Fig. \ref{fig:betamult}.

By Kirchhoff's law and the saturation current relation (by neglecting the Early effect) 
\begin{equation}\EqFontSize
V_{GS,1}=V_{GS,2}+V_{DS,9}
\end{equation}
\begin{equation}\EqFontSize
\sqrt{\frac{2I_{D,1}}{\beta_{n,1}(\sigma)}}+V^{th}_{n,1}=\sqrt{\frac{2I_{D,2}}{\beta_{n,2}(\sigma)}}+V^{th}_{n,2}+V_{DS,9}
\end{equation}
\begin{equation}\EqFontSize
V_{DS,9}=V_{GS,9}-V^{th}_{n,9}\pm \sqrt{(V_{GS,9}-V^{th}_{n,9})^2-\frac{2I_{D,9}}{\beta_{n,9}}} .
\end{equation}

By using the exponential law for the gain, i.e. $\beta=\beta^0\cdot e^{-\pi\sigma}$
\begin{multline}\EqFontSize
\sqrt{\frac{2I_{D,2}}{\beta^0}}\underbrace{\left(\sqrt{\frac{e^{-(\pi_3-\pi_4)\sigma}}{e^{-\pi_1\sigma}}}-\frac{1}{\sqrt{K_{21}e^{-\pi_2\sigma}}}\right)}_{f(K_{21},\sigma)}-\underbrace{\left(V^{th}_{n,2}-V^{th}_{n,1}\right)}_{\Delta V^{th}_n}\\=V_{ov,9}\pm \sqrt{(V_{ov,9})^2-\frac{2I_{D,9}}{\beta_{n9,}}} .
\end{multline}
with $V_{ov,i}=V_{GS,i}-V^{th}_{n,i}$.

The equation is put to the square a first time, giving
 \begin{multline}\EqFontSize\label{eq:beta8}
\frac{2I_{D,2}}{\beta^0}f(K_{21},\sigma)^2-\Delta V^{th}_n\\=(V_{ov,9})^2\pm 2(V_{ov,9})\sqrt{(V_{ov,9})^2-\frac{2I_{D,9}}{\beta_{n,9}}}+V_{ov,9}-\frac{2I_{D,9}}{\beta_{n,9}}.
\end{multline}

with $I_{D,2} = I_{D,9}$ and $\beta^0=\beta^0_{n1}=\frac{\beta^0_9}{K_{91}}$, expression (\ref{eq:beta8}) can be expressed as 

\begin{multline}\EqFontSize
\label{eq:beta9}
\frac{2I_{D,2}}{\beta^0}\underbrace{\left( f(K_{21},\sigma)^2+\frac{1}{K_{91}e^{-\pi_9\sigma}}\right) }_{g(K_{21},K_{91},\sigma)}-\Delta V^{th}_n-(V_{ov,9})^2\\=\pm 2(V_{ov,9})\sqrt{(V_{ov,9})^2-\frac{2I_{D,9}}{\beta_{n,9}}}+V_{ov,9}-\frac{2I_{D,2}}{K_{91}\beta^0e^{-\pi_9\sigma}}.
\end{multline}

We have a second-order equation for $I_{D2}$ by elevating equation (\ref{eq:beta9}) to the square a second time:

\begin{multline}\EqFontSize
4\frac{I_{D,2}^2}{(\beta^0)^2}g(K_{21},K_{91},\sigma)^2+(-\Delta V^{th}_n -2 (V_{ov,9})^2)^2\\+4\frac{I_{D,2}}{\beta^0}g(K_{21},K_{91},\sigma)(-\Delta V^{th}_n -2 (V_{ov,9})^2)\\=4(V_{ov,9})^2((V_{ov,9})^2-\frac{2I_{D,2}}{\beta^0}),
\end{multline}
\begin{multline}\EqFontSize
(I_{D,2})^2\underbrace{\left[\frac{4}{(\beta^0)^2}g(K_{21},K_{91},\sigma)^2\right]}_{A}+I_{D2}\cdot\\\underbrace{\left[\frac{4}{\beta^0}g(K_{21},K_{91},\sigma)(-\Delta V^{th}_n -2 (V_{ov,9})^2)+8\frac{(V_{ov,9})^2}{K_{91}\beta^0e^{-\pi_9\sigma}}\right]}_{B}\\+\underbrace{\left[(-\Delta V^{th}_n -2 (V_{ov,9})^2)^2-4(V_{ov,9})^4\right]}_{C}=0,
\end{multline}

\begin{equation}\EqFontSize
I_{ref}=I_{D2}=\frac{-B\pm\sqrt{B^2-4AC}}{2A}.
\end{equation}

By neglecting the threshold voltage mismatch, $C$ becomes zero. Again, two solutions are obtained with the zero-current one. A start-up circuit is thus needed for this circuit too. The non-zero solution is given by
\begin{align}
I_{D,2}&=\frac{-B\pm\sqrt{B^2}}{2A}\\
&=\frac{-B}{A}\\
&=\frac{-\left[\frac{4}{\beta^0}g(K_{21},K_{91},\sigma)(-2 (V_{ov,9})^2)+8\frac{(V_{ov,9})^2}{K_{91}\beta^0e^{-\pi_9\sigma}}\right]}{\frac{4}{(\beta^0)^2}g(K_{21},K_{91},\sigma)^2}\\
&=2\frac{(V_{ov,9})^2\beta^0 \left(  g(K_{21},K_{91},\sigma)-\frac{1}{K_{91}e^{-\pi_9\sigma}}\right)}{g(K_{21},K_{91},\sigma)^2}\\
&= 2(V_{ov,9})^2\beta^0\frac{\left(\sqrt{\frac{e^{-(\pi_3-\pi_4)\sigma}}{e^{-\pi_1\sigma}}}-\frac{1}{\sqrt{K_{21}e^{-\pi_2\sigma}}}\right)^2}{\left[\left(\sqrt{\frac{e^{-(\pi_3-\pi_4)\sigma}}{e^{-\pi_1\sigma}}}-\frac{1}{\sqrt{K_{21}e^{-\pi_2\sigma}}}\right)^2+\frac{1}{K_{91}e^{-\pi_9\sigma}}\right]^2}.
\end{align}

The reference current can be expressed as
\begin{multline}\EqFontSize
I_{D,2}=2(V_{ov,9})^2\beta_{n,0}\\\cdot \frac{1}{\left[e^{-(\pi_3-\pi_4-\pi_1)\frac{\sigma}{2}}-\frac{e^{\pi_2\frac{\sigma}{2}}}{\sqrt{K_{21}}}+\frac{1}{\left(e^{-(\pi_3-\pi_4-\pi_1)\frac{\sigma}{2}}-\frac{e^{\pi_2\frac{\sigma}{2}}}{\sqrt{K_{21}}}\right)K_{91}e^{-\pi_9\sigma}}\right]^2}.
\end{multline}


\bibliographystyle{elsarticle-num}
\bibliography{biblio.bib}

\begin{thebibliography}{10}
\expandafter\ifx\csname url\endcsname\relax
  \def\url#1{\texttt{#1}}\fi
\expandafter\ifx\csname urlprefix\endcsname\relax\def\urlprefix{URL }\fi
\expandafter\ifx\csname href\endcsname\relax
  \def\href#1#2{#2} \def\path#1{#1}\fi

\bibitem{Cao2015}
H.~Cao, S.~K. Thakar, M.~L. Oseng, C.~M. Nguyen, C.~Jebali, A.~B. Kouki, J.~C.
  Chiao, \href{http://ieeexplore.ieee.org/document/7169510/}{{Development and
  Characterization of a Novel Interdigitated Capacitive Strain Sensor for
  Structural Health Monitoring}}, IEEE Sensors Journal 15~(11) (2015)
  6542--6548.
\newblock \href {http://dx.doi.org/10.1109/JSEN.2015.2461591}
  {\path{doi:10.1109/JSEN.2015.2461591}}.
\newline\urlprefix\url{http://ieeexplore.ieee.org/document/7169510/}

\bibitem{Downey2016}
A.~Downey, S.~Laflamme, F.~Ubertini,
  \href{http://proceedings.spiedigitallibrary.org/proceeding.aspx?doi=10.1117/12.2219301}{{Distributed
  thin film sensor array for damage detection and localization}}, in: J.~P.
  Lynch (Ed.), Proceedings of SPIE - The International Society for Optical
  Engineering, Vol. 9803, Department of Civil, Construction, and Environmental
  Engineering, Iowa State University, Ames, IA, United States, 2016, p. 98030R.
\newblock \href {http://dx.doi.org/10.1117/12.2219301}
  {\path{doi:10.1117/12.2219301}}.
\newline\urlprefix\url{http://proceedings.spiedigitallibrary.org/proceeding.aspx?doi=10.1117/12.2219301}

\bibitem{Chew2017a}
Z.~J. Chew, T.~Ruan, M.~Zhu, M.~Bafleur, J.-M. Dilhac, {Single Piezoelectric
  Transducer as Strain Sensor and Energy Harvester Using Time-Multiplexing
  Operation}, IEEE Transactions on Industrial Electronics 64~(12) (2017)
  9646--9656.
\newblock \href {http://dx.doi.org/10.1109/TIE.2017.2711562}
  {\path{doi:10.1109/TIE.2017.2711562}}.

\bibitem{Yamashita2016}
T.~Yamashita, S.~Takamatsu, H.~Okada, T.~Itoh, T.~Kobayashi, {Ultra-Thin
  Piezoelectric Strain Sensor Array Integrated on a Flexible Printed Circuit
  Involving Transfer Printing Methods}, IEEE Sensors Journal 16~(24) (2016)
  8840--8846.
\newblock \href {http://dx.doi.org/10.1109/JSEN.2016.2578936}
  {\path{doi:10.1109/JSEN.2016.2578936}}.

\bibitem{Liehr2012}
S.~Liehr, K.~Krebber, {Application of quasi-distributed and dynamic length and
  power change measurement using optical frequency domain reflectometry}, IEEE
  Sensors Journal 12~(1) (2012) 237--245.
\newblock \href {http://dx.doi.org/10.1109/JSEN.2011.2157126}
  {\path{doi:10.1109/JSEN.2011.2157126}}.

\bibitem{Guo2016}
Y.~Guo, J.~Kong, H.~Liu, D.~Hu, L.~Qin, {Design and Investigation of a Reusable
  Surface-Mounted Optical Fiber Bragg Grating Strain Sensor}, IEEE Sensors
  Journal 16~(23) (2016) 8456--8462.
\newblock \href {http://dx.doi.org/10.1109/JSEN.2016.2614009}
  {\path{doi:10.1109/JSEN.2016.2614009}}.

\bibitem{Song2006}
K.~Y. Song, Z.~He, K.~Hotate,
  \href{https://www.osapublishing.org/ol/abstract.cfm?uri=ol-31-17-2526}{{Distributed
  strain measurement with millimeter-order spatial resolution based on
  Brillouin optical correlation domain analysis}}, Optics Letters 31~(17)
  (2006) 2526--2528.
\newblock \href {http://dx.doi.org/10.1364/OL.31.002526}
  {\path{doi:10.1364/OL.31.002526}}.
\newline\urlprefix\url{https://www.osapublishing.org/ol/abstract.cfm?uri=ol-31-17-2526}

\bibitem{Fiorillo2018}
A.~Fiorillo, C.~Critello, S.~Pullano,
  \href{https://www.sciencedirect.com/science/article/pii/S0924424718308434}{Theory,
  technology and applications of piezoresistive sensors: A review}, Sensors and
  Actuators A: Physical 281 (2018) 156--175.
\newblock \href {http://dx.doi.org/https://doi.org/10.1016/j.sna.2018.07.006}
  {\path{doi:https://doi.org/10.1016/j.sna.2018.07.006}}.
\newline\urlprefix\url{https://www.sciencedirect.com/science/article/pii/S0924424718308434}

\bibitem{Gridchin1995}
V.~Gridchin, V.~Lubimsky, M.~Sarina,
  \href{https://www.sciencedirect.com/science/article/pii/092442479501013Q}{Piezoresistive
  properties of polysilicon films}, Sensors and Actuators A: Physical 49~(1)
  (1995) 67--72.
\newblock \href
  {http://dx.doi.org/https://doi.org/10.1016/0924-4247(95)01013-Q}
  {\path{doi:https://doi.org/10.1016/0924-4247(95)01013-Q}}.
\newline\urlprefix\url{https://www.sciencedirect.com/science/article/pii/092442479501013Q}

\bibitem{Zymelka2017a}
D.~Zymelka, K.~Togashi, R.~Ohigashi, T.~Yamashita, S.~Takamatsu, T.~Itoh,
  T.~Kobayashi, {Printed strain sensor array for application to structural
  health monitoring}, Smart Materials and Structures 26~(10).
\newblock \href {http://dx.doi.org/10.1088/1361-665X/aa8831}
  {\path{doi:10.1088/1361-665X/aa8831}}.

\bibitem{Pettinato2021}
S.~Pettinato, D.~Barettin, V.~Sedov, V.~Ralchenko, S.~Salvatori,
  \href{https://www.mdpi.com/1996-1944/14/7/1780}{Fabry-perot pressure sensors
  based on polycrystalline diamond membranes}, Materials 14~(7).
\newblock \href {http://dx.doi.org/10.3390/ma14071780}
  {\path{doi:10.3390/ma14071780}}.
\newline\urlprefix\url{https://www.mdpi.com/1996-1944/14/7/1780}

\bibitem{Zhang2019}
J.~X. Zhang, K.~Hoshino,
  \href{https://www.sciencedirect.com/science/article/pii/B9780128148624000065}{Chapter
  6 - mechanical transducers: Cantilevers, acoustic wave sensors, and thermal
  sensors}, in: J.~X. Zhang, K.~Hoshino (Eds.), Molecular Sensors and
  Nanodevices (Second Edition), second edition Edition, Micro and Nano
  Technologies, Academic Press, 2019, pp. 311--412.
\newblock \href
  {http://dx.doi.org/https://doi.org/10.1016/B978-0-12-814862-4.00006-5}
  {\path{doi:https://doi.org/10.1016/B978-0-12-814862-4.00006-5}}.
\newline\urlprefix\url{https://www.sciencedirect.com/science/article/pii/B9780128148624000065}

\bibitem{Jabir2013}
S.~A. Jabir, N.~K. Gupta,
  \href{https://www.sciencedirect.com/science/article/pii/S0263224113000857}{Condition
  monitoring of the strength and stability of civil structures using thick film
  ceramic sensors}, Measurement 46~(7) (2013) 2223--2231.
\newblock \href
  {http://dx.doi.org/https://doi.org/10.1016/j.measurement.2013.03.018}
  {\path{doi:https://doi.org/10.1016/j.measurement.2013.03.018}}.
\newline\urlprefix\url{https://www.sciencedirect.com/science/article/pii/S0263224113000857}

\bibitem{Delhaye2021}
T.~P. Delhaye, N.~André, L.~A. Francis, D.~Flandre, New universal figure of
  merit for embedded si piezoresistive pressure sensors, IEEE Sensors Journal
  21~(1) (2021) 213--221.
\newblock \href {http://dx.doi.org/10.1109/JSEN.2020.3013017}
  {\path{doi:10.1109/JSEN.2020.3013017}}.

\bibitem{Zheng2020}
Q.~Zheng, J.~hun Lee, X.~Shen, X.~Chen, J.-K. Kim,
  \href{https://www.sciencedirect.com/science/article/pii/S1369702119308776}{Graphene-based
  wearable piezoresistive physical sensors}, Materials Today 36 (2020)
  158--179.
\newblock \href
  {http://dx.doi.org/https://doi.org/10.1016/j.mattod.2019.12.004}
  {\path{doi:https://doi.org/10.1016/j.mattod.2019.12.004}}.
\newline\urlprefix\url{https://www.sciencedirect.com/science/article/pii/S1369702119308776}

\bibitem{Ke2018}
K.~Ke, V.~{Solouki Bonab}, D.~Yuan, I.~Manas-Zloczower, {Piezoresistive
  thermoplastic polyurethane nanocomposites with carbon nanostructures}, Carbon
  139 (2018) 52--58.
\newblock \href {http://dx.doi.org/10.1016/j.carbon.2018.06.037}
  {\path{doi:10.1016/j.carbon.2018.06.037}}.

\bibitem{Fisher2012}
G.~Fisher, M.~R. Seacrist, R.~W. Standley, Silicon crystal growth and wafer
  technologies, Proceedings of the IEEE 100~(Special Centennial Issue) (2012)
  1454--1474.
\newblock \href {http://dx.doi.org/10.1109/JPROC.2012.2189786}
  {\path{doi:10.1109/JPROC.2012.2189786}}.

\bibitem{Taroni1970}
A.~Taroni, M.~Prudenziati, G.~Zanarini, Semiconductor sensors:
  Ii—piezoresistive devices, IEEE Transactions on Industrial Electronics and
  Control Instrumentation IECI-17~(6) (1970) 415--421.
\newblock \href {http://dx.doi.org/10.1109/TIECI.1970.230445}
  {\path{doi:10.1109/TIECI.1970.230445}}.

\bibitem{Vittoz1977}
E.~Vittoz, J.~Fellrath, Cmos analog integrated circuits based on weak inversion
  operations, IEEE Journal of Solid-State Circuits 12~(3) (1977) 224--231.
\newblock \href {http://dx.doi.org/10.1109/JSSC.1977.1050882}
  {\path{doi:10.1109/JSSC.1977.1050882}}.

\bibitem{Song1998}
S.~Liu, R.~Baker, Process and temperature performance of a cmos beta-multiplier
  voltage reference, in: 1998 Midwest Symposium on Circuits and Systems (Cat.
  No. 98CB36268), 1998, pp. 33--36.
\newblock \href {http://dx.doi.org/10.1109/MWSCAS.1998.759429}
  {\path{doi:10.1109/MWSCAS.1998.759429}}.

\bibitem{johns2006}
G.~K. Johns, Modeling piezoresistivity in silicon and polysilicon, Journal of
  Applied Engineering Mathematics April 2 (2006) 1--5.

\bibitem{Jastrzebski1988}
Z.~D. Jastrzebski, R.~Komanduri, \href{https://doi.org/10.1115/1.3226051}{{The
  Nature and Properties of Engineering Materials}}, Journal of Engineering
  Materials and Technology 110~(3) (1988) 294.
\newblock \href {http://dx.doi.org/10.1115/1.3226051}
  {\path{doi:10.1115/1.3226051}}.
\newline\urlprefix\url{https://doi.org/10.1115/1.3226051}

\bibitem{Hopcroft2010}
M.~A. Hopcroft, W.~D. Nix, T.~W. Kenny, {What is the Young ' s Modulus of
  Silicon ?}, Journal of Microelectromechanical Systems 19~(2) (2010) 229--238.

\bibitem{Tan2008}
Y.~Tan, X.~Li, L.~Tian, Z.~Yu, {Analytical electron-mobility model for
  arbitrarily stressed silicon}, IEEE Transactions on Electron Devices 55~(6)
  (2008) 1386--1390.
\newblock \href {http://dx.doi.org/10.1109/TED.2008.921074}
  {\path{doi:10.1109/TED.2008.921074}}.

\bibitem{Kanda1982}
Y.~Kanda, \href{http://ieeexplore.ieee.org/document/1482156/}{{A graphical
  representation of the piezoresistance coefficients in silicon}}, IEEE
  Transactions on Electron Devices 29~(1) (1982) 64--70.
\newblock \href {http://dx.doi.org/10.1109/T-ED.1982.20659}
  {\path{doi:10.1109/T-ED.1982.20659}}.
\newline\urlprefix\url{http://ieeexplore.ieee.org/document/1482156/}

\bibitem{Smith1954}
C.~S. Smith,
  \href{https://link.aps.org/doi/10.1103/PhysRev.94.42}{Piezoresistance effect
  in germanium and silicon}, Phys. Rev. 94 (1954) 42--49.
\newblock \href {http://dx.doi.org/10.1103/PhysRev.94.42}
  {\path{doi:10.1103/PhysRev.94.42}}.
\newline\urlprefix\url{https://link.aps.org/doi/10.1103/PhysRev.94.42}

\bibitem{Rue2011}
B.~Rue, B.~Olbrechts, J.-P. Raskin, D.~Flandre,
  \href{http://ieeexplore.ieee.org/document/6081791/}{{A SOI CMOS smart strain
  sensor}}, in: IEEE 2011 International SOI Conference, no.~0, IEEE, 2011, pp.
  1--2.
\newblock \href {http://dx.doi.org/10.1109/SOI.2011.6081791}
  {\path{doi:10.1109/SOI.2011.6081791}}.
\newline\urlprefix\url{http://ieeexplore.ieee.org/document/6081791/}

\bibitem{Jaeger1995}
R.~Jaeger, R.~Ramani, J.~Suhling, Y.~Kang, Cmos stress sensor circuits using
  piezoresistive field-effect transistors (pifets), in: Digest of Technical
  Papers., Symposium on VLSI Circuits., 1995, pp. 43--44.
\newblock \href {http://dx.doi.org/10.1109/VLSIC.1995.520680}
  {\path{doi:10.1109/VLSIC.1995.520680}}.

\bibitem{Jaeger2000}
R.~Jaeger, J.~Suhling, R.~Ramani, A.~Bradley, J.~Xu, Cmos stress sensors on
  [100] silicon, IEEE Journal of Solid-State Circuits 35~(1) (2000) 85--95.
\newblock \href {http://dx.doi.org/10.1109/4.818923}
  {\path{doi:10.1109/4.818923}}.

\bibitem{Tang2003}
S.~Tang, S.~Narendra, V.~De, Temperature and process invariant mos-based
  reference current generation circuits for sub-1v operation, in: Proceedings
  of the 2003 International Symposium on Low Power Electronics and Design,
  2003. ISLPED '03., 2003, pp. 199--204.
\newblock \href {http://dx.doi.org/10.1109/LPE.2003.1231862}
  {\path{doi:10.1109/LPE.2003.1231862}}.

\bibitem{Beaty1992}
R.~Beaty, R.~Jaeger, J.~Suhling, R.~Johnson, R.~Butler, Evaluation of
  piezoresistive coefficient variation in silicon stress sensors using a
  four-point bending test fixture, IEEE Transactions on Components, Hybrids,
  and Manufacturing Technology 15~(5) (1992) 904--914.
\newblock \href {http://dx.doi.org/10.1109/33.180057}
  {\path{doi:10.1109/33.180057}}.

\bibitem{Jeppson2013}
K.~O. Jeppson, {Three- and four-point Hamer-type MOSFET parameter extraction
  methods revisited}, IEEE International Conference on Microelectronic Test
  Structures~(4) (2013) 141--145.
\newblock \href {http://dx.doi.org/10.1109/ICMTS.2013.6528161}
  {\path{doi:10.1109/ICMTS.2013.6528161}}.

\bibitem{Jaeger2018}
R.~C. Jaeger, J.~C. Suhling,
  \href{https://ieeexplore.ieee.org/document/8486881/}{{First and Second Order
  Piezoresistive Characteristics of CMOS FETs: Weak through Strong Inversion}},
  in: 2018 48th European Solid-State Device Research Conference (ESSDERC),
  IEEE, 2018, pp. 126--129.
\newblock \href {http://dx.doi.org/10.1109/ESSDERC.2018.8486881}
  {\path{doi:10.1109/ESSDERC.2018.8486881}}.
\newline\urlprefix\url{https://ieeexplore.ieee.org/document/8486881/}

\bibitem{Wacker2011}
N.~Wacker, H.~Richter, M.-U. Hassan, H.~Rempp, J.~N. Burghartz,
  \href{https://linkinghub.elsevier.com/retrieve/pii/S0038110110004235}{{Compact
  modeling of CMOS transistors under variable uniaxial stress}}, Solid-State
  Electronics 57~(1) (2011) 52--60.
\newblock \href {http://dx.doi.org/10.1016/j.sse.2010.12.003}
  {\path{doi:10.1016/j.sse.2010.12.003}}.
\newline\urlprefix\url{https://linkinghub.elsevier.com/retrieve/pii/S0038110110004235}

\bibitem{Bradley2001}
A.~Bradley, R.~Jaeger, J.~Suhling, K.~O'Connor,
  \href{http://ieeexplore.ieee.org/document/944190/}{{Piezoresistive
  characteristics of short-channel MOSFETs on (100) silicon}}, IEEE
  Transactions on Electron Devices 48~(9) (2001) 2009--2015.
\newblock \href {http://dx.doi.org/10.1109/16.944190}
  {\path{doi:10.1109/16.944190}}.
\newline\urlprefix\url{http://ieeexplore.ieee.org/document/944190/}

\end{thebibliography}

\def\picturelength{3cm}
\vspace{0.5cm}
\parbox[t]{\linewidth}{\par\noindent
\setlength\intextsep{0pt}
\begin{wrapfigure}{L}{\picturelength}
\centering
\includegraphics[width=1.1\linewidth]{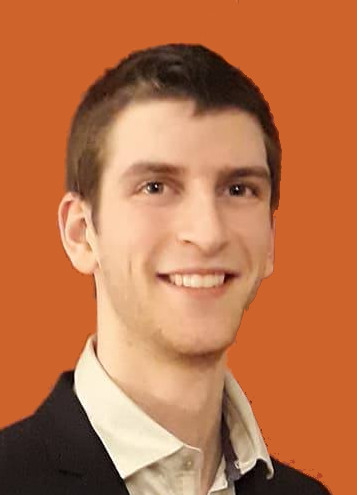}
\end{wrapfigure}
\par\noindent 
{\bf Nicolas Roisin}\
 received the M.S. degree from the University of Louvain, Louvain-la-Neuve, Belgium, in 2019, where he is currently pursuing the Ph.D. degree with the Institute of Information and Communication Technologies, Electronics and Applied Mathematics (ICTEAM). His research topics are oriented towards strained silicon in sensing and optical applications.
}

\vspace{0.5cm}

\parbox[t]{\linewidth}{\par\noindent
\begin{wrapfigure}{L}{\picturelength}
\centering
\vspace{-\baselineskip}
\includegraphics[width=1.1\linewidth]{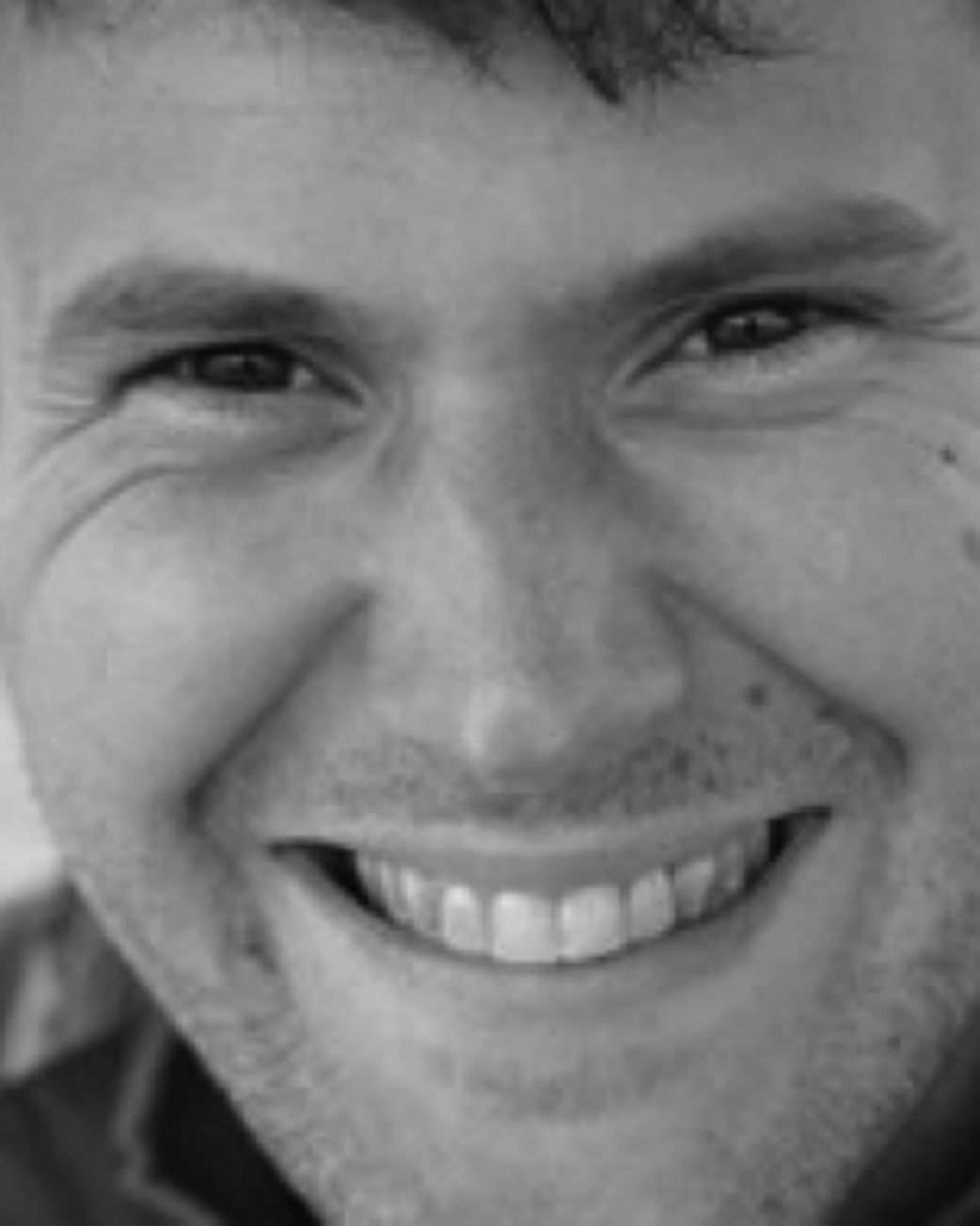}
\end{wrapfigure}
\par\noindent 
{\bf Thibault Delhaye}\
 received the M.S. degree from the University of Louvain, Louvain-la-Neuve, Belgium, in 2015, where he is currently pursuing the Ph.D. degree with the Institute of Information and Communication Technologies, Electronics and Applied Mathematics (ICTEAM). He was a Visiting International Research Student with The University of British Columbia, Canada, in fall 2018. His research topic is on highly sensitive and ultra-low-power MEMS pressure sensor based on SOI technology.
}
\vspace{0.5cm}
\parbox[t]{\linewidth}{\par\noindent
\begin{wrapfigure}{L}{\picturelength}
\centering
\vspace{-\baselineskip}
\includegraphics[width=1.05\linewidth]{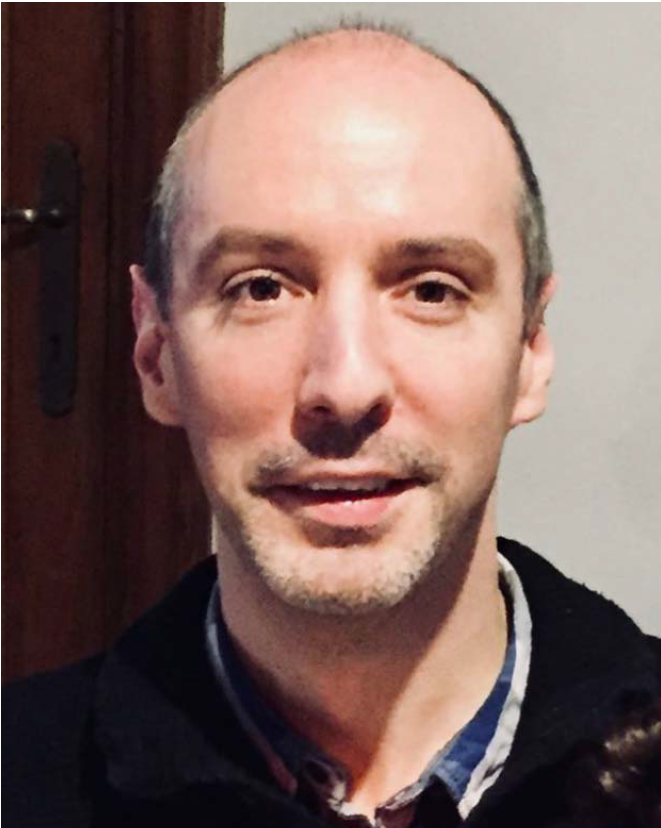}
\end{wrapfigure}
\par\noindent 
{\bf Nicolas André}\
 received the M.S. degree in electrical engineering from the Louvain School of Engineering, Université Catholique de Louvain (UCL), Louvain-la-Neuve, Belgium, in 2004, and the Ph.D. degree in applied sciences in the field of microelectromechanical systems (MEMS) co-integration from UCL in 2011. From 2011 to 2012, he was with UdeS, Sherbrooke, Canada, as a Postdoctoral Researcher on bio-inspired methods to improve the LED efficiency. He has coauthored more than 100 research articles in international journals and holds two patents. He was a team member in several Walloon, FEDER, and EU projects as STARFLO+, FEDER MINATIS, and MICRO+. His expertise is about microfabrication and sensors (flow, humidity, pressure, and light) integrated with SOI CMOS Circuits.
}
\vspace{0.5cm}
\parbox[t]{\linewidth}{\par\noindent
\begin{wrapfigure}{L}{\picturelength}
\centering
\vspace{-\baselineskip}
\includegraphics[width=1.1\linewidth]{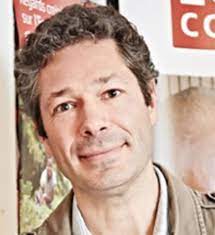}
\end{wrapfigure}
\par\noindent 
{\bf Jean-Pierre Raskin}\
 (M’97–SM’06) received the M.S. and Ph.D. degrees in applied sciences from the Université catholique de Louvain, Louvain-la-Neuve, Belgium, in 1994 and 1997, respectively. He was a Visiting Professor with Newcastle University, Newcastle upon Tyne, U.K., from 2009 to 2010. His research interests are the modeling, wideband characterization, and fabrication of advanced SOI MOSFETs as well as micro and nanofabrication of MEMS/NEMS sensors and actuators, including the extraction of intrinsic material properties at nanometer scale.
}

\vspace{0.5cm}

\parbox[t]{\linewidth}{\par\noindent
\begin{wrapfigure}{L}{\picturelength}
\centering
\vspace{-\baselineskip}
\includegraphics[width=1.1\linewidth]{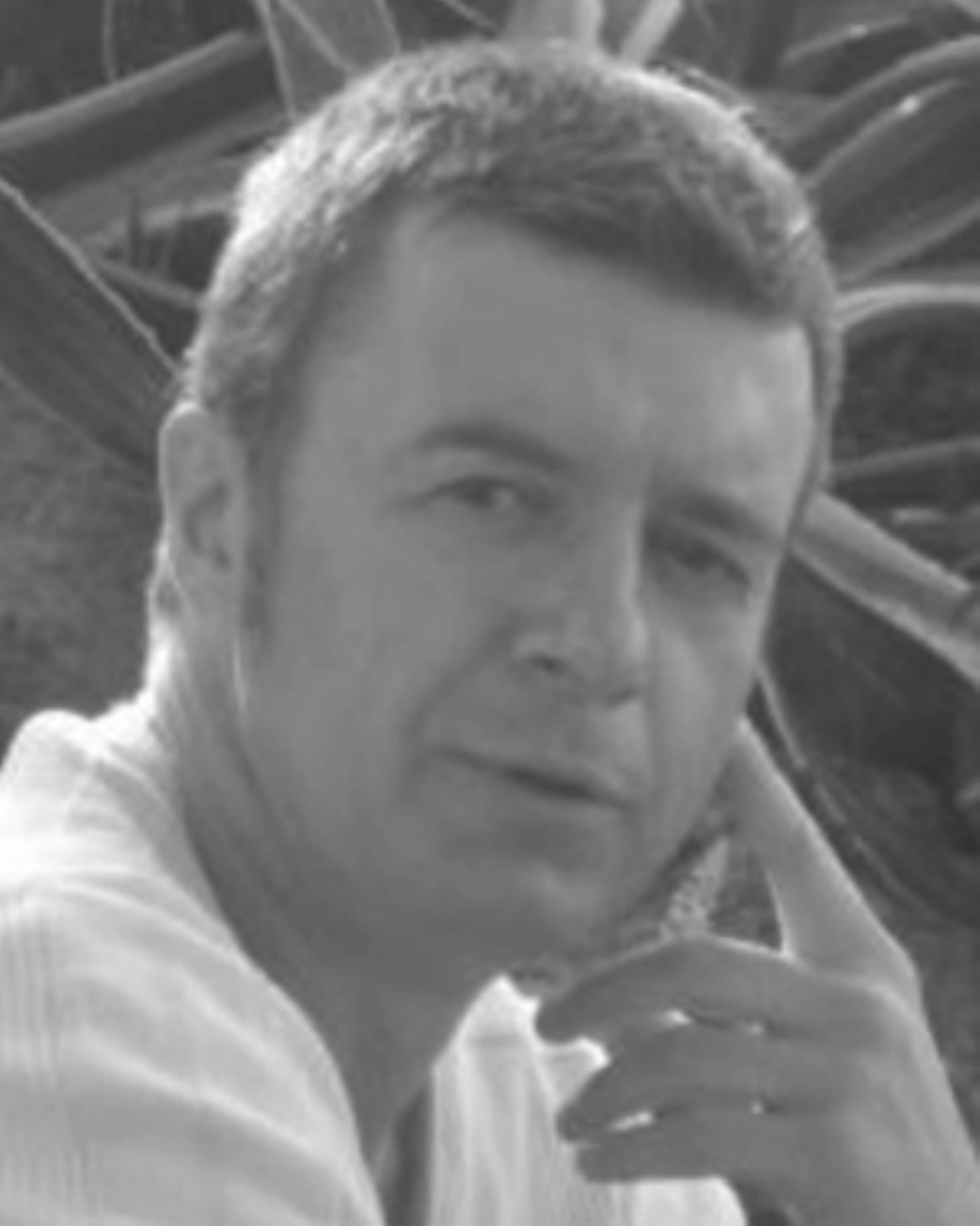}
\end{wrapfigure}
\par\noindent 
{\bf Denis Flandre}\ received the M.S. degree in electrical engineering, and the Ph.D. and Research Habilitation degrees from the Université catholique de Louvain (UCL), Louvain-la-Neuve, Belgium, in 1986, 1990 and 1999, respectively. His doctoral research was on the modeling of Silicon-on-Insulator (SOI) MOS devices for characterization and circuit simulation, his Postdoctoral thesis on a systematic and automated synthesis methodology for MOS analog circuits. Since 2001, he has been a full-time Professor with UCL. He has participated or coordinated numerous research projects funded by regional and European institutions. He has organized or lectured many short courses on SOI technology, devices and circuits in universities, industrial companies, and conferences. He has authored or coauthored more than 1000 technical papers or conference contributions. He is a Co-Inventor of 11 patents. He is currently involved in the research and development of SOI MOS devices, digital and analog circuits, as well as sensors and MEMS, for special applications, more specifically high-speed, low-voltage low-power, microwave, biomedical, radiation-hardened, and high-temperature electronics and microsystems.}

\end{document}